\documentclass[a4paper, 11pt, oneside, notitlepage]{article}
\usepackage[T1]{fontenc}
\usepackage [english]{babel}
\usepackage{natbib} %Serve per la bibliografia

\usepackage{amsmath,amssymb}
\usepackage{tabularx}
\usepackage{latexsym}
\usepackage{amsfonts}
\usepackage{mathtools} %to be used to define =^H
\usepackage{bm}%bold greek letters
\usepackage{dsfont}%indicator operator
\usepackage{bbm}%indicator operator

\usepackage{amsthm}
\usepackage{amscd}
\usepackage[inline]{enumitem}%easy to customize than \enumerate

\usepackage{multirow}
\usepackage{graphicx,graphics,rotating}
\usepackage[graphicx]{realboxes} %rotate tables
\usepackage{epstopdf} % for eps files
\usepackage[font=scriptsize,labelfont=bf,labelsep=period,textfont=it]{caption}

\usepackage{mathrsfs}
\usepackage[a4paper,top=2.5cm,bottom=2.5cm,left=2cm,right=2cm,height rounded,bindingoffset=0.5cm]{geometry}

\usepackage{etoolbox}
\usepackage{xpatch}  
\usepackage[gen]{eurosym}
\usepackage{lscape}
\usepackage[title,titletoc]{appendix}

\usepackage{sgame}
\usepackage{setspace}
\usepackage[font={small,sl}]{caption}
\usepackage{booktabs,caption} %to be used w/ \captionsetup
\usepackage[para,online,flushleft]{threeparttable}	%to align notes under table

\usepackage{blindtext}
\usepackage{fancyhdr}
\usepackage{setspace}
\usepackage[explicit]{titlesec}
\usepackage[normalem]{ulem}
\usepackage{lipsum}
\usepackage{mwe}  
\usepackage{verbatim}
\usepackage{booktabs,array}  
\usepackage{listings}
\usepackage{subfig}
\usepackage[colorlinks=true,citecolor=blue,linkcolor = blue,urlcolor  = blue]{hyperref}
\usepackage{xcolor}

\usepackage{booktabs}
\usepackage[hang,flushmargin]{footmisc}

\makeatletter

\def\citeapos#1{\citeauthor{#1}'s (\citeyear{#1})}%\citeapos

\title{\Large \bf Robust Inference in Panel Data Models: \\
{\large \bf Some Effects of Heteroskedasticity and Leveraged Data in Small Samples}}
\author{Annalivia Polselli\thanks{Institute of Analytics and Data Science (IADS) and Centre for Micro-Social Change (MiSoc), University of Essex. Email: \href{annalivia.polselli@essex.ac.uk}{annalivia.polselli@essex.ac.uk}. This is a preliminary revised version of the first chapter of my doctoral thesis, available at \url{https://repository.essex.ac.uk/34481/}. This project was supported by the South East Network for Social Sciences (SeNSS) as part of the Doctoral Training Partnership (award ES/P00072X/1/2128216).}}
\date{\today}

\begin{document}

\maketitle

%figcaotion
\newcommand\fnote[1]{\captionsetup{font=small}\caption*{#1}}

%colors
\newcommand{\red}[1]{\textcolor{red}{#1}}
\newcommand{\blue}[1]{\textcolor{blue}{#1}}
\newcommand{\orange}[1]{\textcolor{Orange}{#1}}
\newcommand{\lightb}[1]{\textcolor{Cyan}{#1}}
\newcommand{\gr}{\cellcolor{YellowGreen}} 
\newcommand{\bad}{\cellcolor{OrangeRed}} 
\newcommand{\yellow}{\cellcolor{Yellow}} 
%maths symbols
\newcommand{\distas}[1]{\mathbin{\overset{#1}{\kern\z@\sim}}}%
\newsavebox{\mybox}\newsavebox{\mysim}
\newcommand{\distras}[1]{%
  \savebox{\mybox}{\hbox{\kern3pt$\scriptstyle#1$\kern3pt}}%
  \savebox{\mysim}{\hbox{$\sim$}}%
  \mathbin{\overset{#1}{\kern\z@\resizebox{\wd\mybox}{\ht\mysim}{$\sim$}}}%
}
%\newcommand\myeq{\stackrel{\mathclap{\normalfont{H}}}{=}} %write on equal sign

%betas
\newcommand{\bfe}{\widehat{\bm{\beta}}}
\newcommand{\bfen}{\widehat{\bm{\beta}}_N}
\newcommand{\tbfen}{\widehat{\widetilde{\bm{\beta}}}_N}
\newcommand{\bi}{\widehat{\bm{\beta}}_{(i)}}
\newcommand{\bij}{\widehat{\bm{\beta}}_{(i,j)}}
\newcommand{\bj}{\widehat{\bm{\beta}}_{(j)}}
\newcommand{\bmean}{\bar{\bm{\beta}}}
\newcommand{\bbeta}{\bm{\beta}}
%x
\newcommand{\x}{\mathbf{x}_{it}}
\newcommand{\w}{\mathbf{w}_{it}}
\newcommand{\xit}{\widetilde{\mathbf{x}}_{it}}
\newcommand{\xis}{\widetilde{\mathbf{x}}_{is}}
\newcommand{\xii}{\mathbf{X}_i}
\newcommand{\xij}{\widetilde{\mathbf{X}}_j}
\newcommand{\Xii}{\widetilde{\mathbf{X}}_i}
\newcommand{\Xjj}{\widetilde{\mathbf{X}}_j}
\newcommand{\Xll}{\widetilde{\mathbf{X}}_l}
\newcommand{\Xj}{\widetilde{\mathbf{X}}_{i(j)}}
\newcommand{\Xij}{\widetilde{\mathbf{X}}_{ij}}
\newcommand{\X}{\widetilde{\mathbf{X}}}
\newcommand{\XX}{\mathbf{X}}
\newcommand{\Xmodel}{\mathbf{X}}
\newcommand{\SSxx}{\mathbf{S}_{xx}}
\newcommand{\SSXX}{\mathbf{S}_{XX}}
\newcommand{\tSSxx}{\widetilde{\mathbf{S}}_{xx}}
\newcommand{\Sxx}{\mathbf{S}_N}
\newcommand{\Sxixi}{\mathbf{S}_{iN}}
\newcommand{\tSxx}{\widehat{\widetilde{\mathbf{S}}}_N}
\newcommand{\sxy}{\hat{\mathbf{s}}}
\newcommand{\bSn}{\overline{\mathbf{S}}_N}
\newcommand{\bVn}{\overline{\mathbf{V}}_N}
\newcommand{\tbSn}{\widetilde{\overline{\mathbf{S}}}_N}
\newcommand{\tbVn}{\widetilde{\overline{\mathbf{V}}}_N}
\newcommand{\Vi}{\mathbf{V}_i}
\newcommand{\Vj}{\mathbf{V}_j}
%y

\newcommand{\y}{y_{it}}
\newcommand{\yii}{\mathbf{y}_i}
\newcommand{\yit}{\widetilde{y}_{it}}
\newcommand{\yi}{\widetilde{\mathbf{y}}_i}
\newcommand{\yj}{\widetilde{\mathbf{y}}_j}
\newcommand{\yij}{\widetilde{\mathbf{y}}_{ij}}
\newcommand{\Y}{\widetilde{\mathbf{Y}}}
\newcommand{\Ymodel}{\mathbf{Y}}

%h
\newcommand{\hit}{h_{it}}
\newcommand{\histar}{h^*_{i}}
\newcommand{\hitt}{h_{itt}}
\newcommand{\hits}{h_{its}}
\newcommand{\bhtt}{\overline{h}_{tt}}
\newcommand{\bhvec}{\overline{\mathbf{h}}}
\newcommand{\Hii}{\mathbf{H}^*_{ii}}
\newcommand{\Hi}{\mathbf{H}_i}
\newcommand{\hi}{\mathbf{h}_{i}}
\newcommand{\bHtt}{\mathbf{\overline{h}}_{tt}}

\newcommand{\bh}{\overline{\overline{h}}}
\newcommand{\Bh}{\overline{\overline{\mathbf{h}}}}

\newcommand{\bH}{\mathbf{\overline{h}}}
\newcommand{\Hj}{\mathbf{H}_j}
\newcommand{\Hij}{\mathbf{H}_{ij}}
\newcommand{\Hji}{\mathbf{H}_{ji}}
\newcommand{\blockH}{\mathbf{\overline{H}}_{ij}}
\newcommand{\HH}{\mathbf{H}}
\newcommand{\tHii}{\widetilde{\mathbf{H}}^*_{ii}}
\newcommand{\tHi}{\widetilde{\mathbf{H}}^*_{i}}

%Mi
\newcommand{\M}{\mathbf{M}}
\newcommand{\Mi}{\mathbf{M}_i}
\newcommand{\Mj}{\mathbf{M}_j}
\newcommand{\Mij}{\mathbf{M}_{ij}}
\newcommand{\Mji}{\mathbf{M}_{ji}}
\newcommand{\blockM}{\mathbf{\overline{M}}_{ij}}
\newcommand{\xMi}{\mathbf{M}_i^{\xi_i}}
\newcommand{\dMi}{\mathbf{M}_i^{\delta_i}}

%u
\newcommand{\error}{u_{it}}
\newcommand{\errori}{\mathbf{u}_i}
\newcommand{\stdres}{\mathbf{r}_i}
\newcommand{\errorj}{\widetilde{\mathbf{u}}_j}
\newcommand{\errorl}{\widetilde{\mathbf{u}}_l}
\newcommand{\uit}{\widehat{u}_{it}}
\newcommand{\tildeuit}{\widetilde{u}_{it}}
\newcommand{\tuit}{\widehat{\widetilde{u}}_{it}}
\newcommand{\uii}{\widetilde{\mathbf{u}}_i}
\newcommand{\ujj}{\widetilde{\mathbf{u}}_j}
\newcommand{\ull}{\widetilde{\mathbf{u}}_l}
\newcommand{\tui}{\widehat{\widetilde{\mathbf{u}}}_i}
\newcommand{\ui}{\widehat{\mathbf{u}}_i}
\newcommand{\uitt}{\widehat{\widetilde{\mathbf{u}}}_i}
\newcommand{\resi}{\widehat{\mathbf{u}}_i}
\newcommand{\resj}{\widehat{\mathbf{u}}_j}
\newcommand{\resl}{\widehat{\mathbf{u}}_l}
\newcommand{\resij}{\widehat{\mathbf{u}}_{ij}}
\newcommand{\U}{\widehat{\widetilde{\mathbf{U}}}}
\newcommand{\UU}{\widehat{\mathbf{U}}}
\newcommand{\Ut}{\widetilde{\mathbf{U}}}
\newcommand{\Error}{\mathbf{U}}
\newcommand{\vi}{\widehat{\mathbf{v}}_i}

\newcommand{\uitthat}{\widehat{u}_{itt}}
\newcommand{\uittstar}{\widehat{u}^*_{itt}}
\newcommand{\uistar}{\widehat{u}^*_{i}}
\newcommand{\Uistar}{\widehat{\mathbf{u}}^*_{i}}
\newcommand{\uitstar}{\widehat{u}^*_{it}}
\newcommand{\uith}{\widehat{\widetilde{u}}^*_{it}}
\newcommand{\tuih}{\widehat{\widetilde{\mathbf{u}}}_i^*}
\newcommand{\uih}{\widehat{\mathbf{u}}_i^*}
\newcommand{\uitth}{\widehat{\widetilde{\mathbf{u}}}_i^*}
\newcommand{\Uh}{\widehat{\widetilde{\mathbf{U}}}{^*}}
\newcommand{\uh}{\widehat{\mathbf{U}}{^*}}

\newcommand{\epsit}{\epsilon_{it}}
\newcommand{\epsi}{\bm{\epsilon}_i}
\newcommand{\tepsi}{\widetilde{\bm{\epsilon}}_i}
\newcommand{\sigmaeps}{\sigma^2_{\epsilon,it}}
\newcommand{\hsigmaeps}{\widehat{\sigma}^2_{\epsilon,it}}
\newcommand{\tepsit}{\widetilde{\epsilon}_{it}}

%letters
\newcommand{\A}{\mathrm{A}}
\newcommand{\Amat}{\mathrm{A}(\Xii)}
\newcommand{\Ajmat}{\mathrm{A}(\Xjj)}
\newcommand{\ANmat}{\mathrm{A}_N(\Xii)}
\newcommand{\B}{\mathrm{B}}
\newcommand{\Bmat}{\mathrm{B}(\Xii)}
\newcommand{\Bjmat}{\mathrm{B}(\Xjj)}
\newcommand{\BNmat}{\mathrm{B}_N(\Xii)}
\newcommand{\Dmat}{\mathrm{D}_1\big(\Xll\big)}
\newcommand{\DDdot}{\mathrm{D}_1(.)}
\newcommand{\Ddmat}{\mathrm{D}_2\big(\Xll,\ull,\bbeta\big)}
\newcommand{\Dddot}{\mathrm{D}_2(.)}
\newcommand{\C}{\mathrm{C}}
\newcommand{\Ci}{\mathrm{C}_{ii}(\bfe)}
\newcommand{\Cij}{\mathrm{C}_{ij}(\bfe)}
\newcommand{\cCij}{\mathrm{C}_{i(j)}(\bfe)}
\newcommand{\D}{\mathrm{D}}
\newcommand{\Q}{\mathrm{Q}}
\newcommand{\Kij}{\mathrm{K}_{j|i}}
\newcommand{\MSij}{\mathrm{M}_{i(j)}}
\newcommand{\F}{\mathrm{F}}

\newcommand{\NID}{\mathrm{NID}}
\newcommand{\Vjk}{\widehat{\mathrm{AVar}(\bfe)}_{jk}}
\newcommand{\V}{\widehat{\mathrm{AVar}(\tbfe)}}
\newcommand{\AVar}{\widehat{\mathrm{AVar}(\bfe_{N,T,r})}}
\newcommand{\avar}{\widehat{\mathrm{AVar}(\bfe)}}
\newcommand{\aavar}{\mathrm{Avar}_{\bbeta}}
\newcommand{\Var}{\mathrm{Var}}
\newcommand{\Avar}{\mathrm{Avar}}
\newcommand{\Cov}{\mathrm{Cov}}
\newcommand{\mustar}{\bm{\mu}^*}
\newcommand{\Dd}{\mathbf{D}}
\newcommand{\Dh}{\widehat{\mathbf{D}}}
\newcommand{\md}{\mathbf{M}}
\newcommand{\lt}{\mathbf{L}}
\newcommand{\bfj}{\mathbf{j}}
\newcommand{\iotat}{\bm{\iota}}
\newcommand{\Vv}{\mathbf{V}}
\newcommand{\Z}{\mathbf{Z}}
\newcommand{\tZ}{\widetilde{\mathbf{Z}}}
\newcommand{\tVv}{\widetilde{\mathbf{V}}_i}
\newcommand{\Ga}{\bm{\Gamma}\big(\bfe,\bi\big)}

\newcommand{\Ti}{T_i}
\newcommand{\Vh}{\widehat{\mathbf{V}}_N}
\newcommand{\Vhh}{\widehat{\widehat{\mathbf{V}}}_N}
\newcommand{\vh}{\widehat{\mathbf{V}}}
\newcommand{\tVh}{\widehat{\widetilde{\mathbf{V}}}}
\newcommand{\inv}{^{-1}}
\newcommand{\invsqrt}{^{-1/2}}
\newcommand{\asqrt}{^{1/2}}

\newcommand{\plim}{\mathrm{plim}}
\newcommand{\N}{\mathcal{N}}
\newcommand{\uniform}{\mathcal{U}}
\newcommand{\Cc}{\mathcal{C}}
\newcommand{\E}{\mathbb{E}}
\newcommand{\R}{\mathbb{R}}  %uncomment
\newcommand{\W}{\mathrm{W}}
\newcommand{\Sgma}{\bm{\Sigma}_i}
\newcommand{\Sgmaj}{\bm{\Sigma}_j}
\newcommand{\Sgmal}{\bm{\Sigma}_l}
\newcommand{\Ssgma}{\bm{\Sigma}}
\newcommand{\Sgmahat}{\widehat{\bm{\Sigma}}_i}
\newcommand{\Sigmatilde}{\widetilde{\bm{\Sigma}}_i}
\newcommand{\bSgma}{\overline{\bm{\Sigma}}_N}

\renewcommand{\th}{^{th}}%%n^{th}
\newcommand{\tr}{\mathrm{tr}}
\newcommand{\diag}{\mathrm{diag}}
\renewcommand\qedsymbol{$\blacksquare$}
\newcommand{\bb}{\big\|}
\newcommand{\Bb}{\Big\|}
%identity
\newcommand{\one}{\mathds{1}}
\newcommand{\I}{\mathbf{I}}
\newcommand{\Ii}{\mathbf{I}_i}
\newcommand{\Ij}{\mathbf{I}_j}
\newcommand{\Iij}{\mathbf{I}_{ij}}

%zeros
\newcommand{\zero}{\bm{0}}
\newcommand{\Oij}{\bm{0}_{ij}}
\newcommand{\Oji}{\bm{0}_{ji}}
\newcommand{\hlambda}{\widehat{\lambda}_{it}}

%greek
\newcommand{\bmu}{\bm{\mu}}
\newcommand{\bsigma}{\bm{\sigma}}

%%%%%%%%%%%%%%%%%%%%%%%%%
%%%%%%%%%%%%%%%%%%%%%%%%%%%%%
\begin{abstract}
	%\doublespacing
	\noindent 
With the violation of the assumption of homoskedasticity, least squares estimators of the variance become inefficient and statistical inference conducted with invalid standard errors leads to misleading rejection rates. Despite a vast cross-sectional literature on the downward bias of robust standard errors, the problem is not extensively covered in the panel data framework. We investigate the consequences of the simultaneous presence of small sample size, heteroskedasticity and data points that exhibit extreme values in the covariates (`good leverage points') on the statistical inference. Focusing on one-way linear panel data models, we examine asymptotic and finite sample properties of a battery of heteroskedasticity-consistent estimators using Monte Carlo simulations. We also propose a hybrid estimator of the variance-covariance matrix.  Results show that conventional standard errors are always dominated by more conservative estimators of the variance, especially in small samples. In addition, all types of HC standard errors have excellent performances in terms of size and power tests under homoskedasticity.\\
	

\end{abstract}
\section{Introduction}
%%Framework+Motivation
\par When the assumption of homoskedasticity is violated and the disturbances show non-constant variance (within the cross-sectional timention, or time dimension, or both), least squares (LS) estimators are no longer efficient. Consequently,  standard errors based on the incorrect assumption of homoskedastic disturbances lead to misleading statistical inferences.   A common practice is to account for heteroskedasticity with  robust standard errors when estimating the model.  The Eicker-Huber-White (EHW) estimator \citep{eicker1967,huber1967,white1980}  has become the norm to account for any degree of heteroskedasticity in the cross-sectional environment. Its counterpart for the panel data is the \mbox{\citeapos{arellano1987}} formula.
The presence of data points that exhibit extreme values in the covariates -- i.e., \emph{good leverage points} --
makes the EHW estimator systematically downward biased leading to liberal statistical inferences  \citep{long2000,godfrey2006,hayes2007,mackinnon2013,csimcsek2016}. The bias is severe when the cross-sectional sample size is sufficiently small  (e.g., with less than 250 units in the sample), and persists even in large samples \citep{mackinnon1985, chesher1987,silva2001,verardi2009}.  While much discussion has involved the cross-sectional framework, little has been investigated for panel data, despite similar issues of \citeapos{arellano1987} standard errors.\footnote{To the best of our knowledge, there are only two available studies for panel data. \citet{kezdi2003} compares the finite sample properties of a series of estimators of the variance-covariance matrix with an without serial correlation in the error term in large-N and small-T panels.   \citet{hansen2007} derive the asymptotic properties of the conventional estimator of the variance-covariance matrix and studies its finite sample behaviour under heteroskedasticity in the cases where both $N,T$ jointly go to infinity, and where either $N$ or $T$ goes to infinity holding the other dimension fixed. Extensions of a class of HC-based estimators to linear panel data mode ls has been conducted by  \citet{cattaneo2018}  in  high dimensional literature.}.

%In some economic fields, researchers often face a data structure with rather small number of units, heteroskedasticity, and unusual data points. Such data structure is common in macroeconomic country-level analyses, and applied experimental and behavioural studies. 

In this paper, we investigate the consequences of the simultaneous presence of small sample size, good leveraged data, and heteroskedastic disturbances on the validity of the statistical inference in linear panel data models.  We formalise panel versions of \citeapos{mackinnon1985} and \citeapos{davidson1993} estimators,  and propose a new hybrid estimator, $PHC6$, that  penalises only units with high leverage in the covariates. We derive the asymptotic distributions of this battery of estimators, and analyse their finite sample properties with Monte Carlo (MC) simulations in terms of proportional bias, rejection probability (or empirical size), root mean squared error, and adjusted power. The analysis is conducted across different panel sample sizes  and  degrees of heteroskedasticity. Units are randomly contaminated with good leverage points. While we treat homoskedasticity as a special case, heteroskedasticity is assumed to be a core component of the correct regression specification.  

We find that under  heteroskedasticity and with good leveraged data test statistics obtained with \citeapos{arellano1987} standard errors are, as expected, over-sized, upward biased, and with low power, especially when the panel size is smaller than 2,500 observations. Test statistics calculated with PHC6 formula  mimic the behaviour of those based on jackknife standard errors in terms of bias, empirical size and adjusted power test, converging  to the same rates  as the sample size increases. The panel version of \citeapos{mackinnon1985} estimator shows similar patterns but with different magnitudes. 
Under homoskedasticity and with good leveraged data, all estimators  have good performances in terms of proportional bias, rejection probabilities, and  adjusted power, suggesting that the heteroskedasticity correction should be used. A similar result was found in  \citet{mackinnon1985} and \citet{long2000} for cross-sectional models who claimed that jackknife-type standard errors might enhance inference even with small degrees of heteroskedasticity. 

We focus on small sample sizes for a two reasons. First, the cross-sectional HC literature has extensively discussed the finite sample bias of the EHW estimator in the presence of leverage points, and we want to document the behaviour of \citeapos{arellano1987} estimator under the same circumstances. Second, the nature of the research and/or data availability may force the investigator to deal with a reduced number of observations in the dataset. %Furthermore, we are interested in the additional issue posed by the presence of good leveraged data in this setting because they may be carried over the full history of a unit and, hence, contribute to exacerbate the effect on the estimates of the variance.

Despite the remarkable methodological contribution in the cross-sectional HC literature, HC-type estimators\footnote{HC-type estimators include: HC2 by \citet{horn1975}, HC3 by \citet{mackinnon1985}, HC$jk$ by \citet{davidson1993}, HC4 by \citet{cribari2004}, HC5 by \citet{cribari2007}, and HC4m by \citet{cribari2011}.} have not found much application in practice, although by construction they alleviate the effect of  leveraged data being less sensitive to  anomalous cases \citep{hinkley1977}. This study contributes to the HC literature by creating a link between cross-sectional and panel HC estimators of the sampling variance. We provide the formulae and derive the distribution of a selected group of variance-covariance estimators to panel data. We document the downward bias of conventional robust standard errors under certain circumstances and provide alternative solutions to obtain more reliable statistical inferences. This study provides simulation evidence that these estimators outperform the conventional cluster-robust standard errors under specific circumstances and should be used in linear panel data models.

The rest of the paper is structured as follows. Section \ref{sec:model} introduces the static linear panel data model and its assumptions, and the asymptotic properties of the \emph{within-group} estimator. In Section~\ref{sec:asyvar}, we discuss the estimation of the variance-covariance matrix, formalise HC estimators for panel data and propose a new estimator.  Section~\ref{sec:mc_sim} shows the MC simulation design and discusses the simulation results. In Section~\ref{sec:simulres}, we examine the performances of the four estimators in terms of their proportional bias, empirical size, adjusted power, and mean squared errors. Section~\ref{sec:conclusion} concludes.

\section{The Model and Estimator}\label{sec:model}
\subsection{Model and Assumptions}
%new version
\noindent Consider the static linear panel regression model with one-way error component
\begin{equation}\label{eq:fe}  
\y =  \x' \bm{\beta} + \alpha_i + \error, \hspace{5pt}  i \in \mathcal{I}=\{1,\dots,N\} \hspace{3pt}  \text{and} \hspace{3pt}  t\in \mathcal{T}=\{1,\dots,T\}
\end{equation}

\noindent where $\y$ is the response variable for the cross-sectional unit $i$ at time period $t$; $\x$ is a $k\times1$ vector of time-varying inputs, $\bm{\beta}$ is a $k\times1$ vector of parameters of interest; $\alpha_i$ is the individual-specific unobserved heterogeneity (or \emph{fixed effects}); and $\error$ is a stochastic error component.

%%%%
\par Stacking observations for $t$, model \eqref{eq:fe} at the level of the observation becomes
 \begin{equation}\label{eq:fe_i}   
\yii =   \xii\bm{\beta} +\bm{\alpha}_i+ \errori,  \,\,\, \text{for all}  \,\,\, i= 1,\dots, N, 
\end{equation}
\noindent where $\yii$ is $T\times1$ vector of outcomes; $\xii$ is a $T\times k$ matrix of  time-varying regressors; $\bm{\alpha}_i=\alpha_i\iotat$ is a $T\times1$ vector of individual fixed effects, and $\iotat$ is a vector of ones of order $T$; and $\errori$ is a $T\times1$ vector of one-way error component.  
%This  formulation allows for  models that either include a random  individual-specific component, independent of the observables, in the compound error \footnote{In models with random effects, the one way error term is defined as the sum of two vector components $\epsi$ and $\bm{\alpha}_i$, where individual fixed-effect is $\bm{\alpha}_i=\iotat\alpha_i$, and $\iotat$ is a vector of ones}
%\par Under the assumption \mbox{$\E(\bm{\alpha}_i|\xii)=h(\xii)$}, the parameter of interest $\bm{\beta}$ can be consistently estimated applying an appropriate transformation to the data, i.e., \emph{time-demeaning} or \emph{first-differencing} the original model. 
The fixed effects $\bm{\alpha}_i$ in Equation~\eqref{eq:fe_i} are removed to consistently estimating the parameter of interest $\bm{\beta}$ by applying an appropriate transformation of the original data, i.e., the \emph{time-demeaning} or \emph{first-differencing} procedure, because it might be the case that \mbox{$\E(\bm{\alpha}_i|\xii)=h(\xii)$}. For the rest of the discussion, we focus on the first approach when applied to Equation~\eqref{eq:fe_i}. The \emph{time-demeaning}  data transformation delivers a consistent estimator of $\bm{\beta}$ even when the regressor is correlated with the unobserved heterogeneity $\alpha_i$, but is less efficient than the First-Difference (FD) transformation with errors that are not identically distributed.

The estimating equation becomes
%%%%%
 \begin{equation}\label{eq:fe_wg_i}   
\yi =  \Xii\bm{\beta} + \uii, \,\,\, \text{for all}  \,\,\, i= 1,\dots, N,  
\end{equation}
\noindent where $\yi=(\I_T-\, T\inv\iotat\iotat')\yii$ is $T\times1$; $\Xii=(\I_T-\, T\inv\iotat\iotat')\xii$ is $T\times k$; and $\uii=(\I_T-\, T\inv\iotat\iotat')\errori$ is $T\times1$. Note that $(\I_T-\, T\inv\iotat\iotat')\bm{\alpha}_i=\zero$ as $T\inv \iotat\iotat' \bm{\alpha}_i=\bm{\alpha}_i$.  The  within-group estimator is the Pooled OLS estimator of Equation~\eqref{eq:fe_wg_i}. 
%\par Henceforth, we state the model assumptions and estimator's properties in terms of the variables $\yii$, $\xii$ and $\errori$ that in models with unobserved heterogeneity can be interpreted as random variables whose data transformation removed the presence of the individual-specific fixed effects from the equation, as mentioned above. 

The model assumptions are as follows
%%%
\begin{enumerate}[label=\sc{asm.}{\arabic*}, leftmargin=1.7\parindent,  rightmargin=.5\parindent]
\item (\emph{data-generating process}): \label{item:dgp} %%
 	\begin{enumerate}[label={\roman*}]
	\item  (\emph{independent variables}): $\{\xii\}$ is an independent and identically distributed (\emph{iid}) sequence of random variables, for all $\, i = 1, \dots, N$; \label{item:dgp_x}
	\item  (\emph{disturbances}): $\{\errori\}$ is an independent  but not identically distributed (\emph{inid}) sequence of random error terms, for all $\, i = 1, \dots, N$.  \label{item:dgp_u}
	\end{enumerate}

\item (\emph{on the relation of $\Xii$ and $\uii$}):\label{item:xu}
 	\begin{enumerate}[label={\roman*}]
	\item (\emph{strong exogeneity}): $\E\bigl(\uii|\Xii\bigr)=0$, for all $i= 1,\dots, N$; \label{item:exogeneity}
	\item  (\emph{heteroskedasticity}): $\bSgma=N\inv \sum_{i=1}^N\Sgma\to\Ssgma$, where the  matrix of the heteroskedastic disturbances $\Sgma=\E\bigl(\uii\uii'|\Xii\bigr)= \mathrm{diag}\bigl\{\sigma_{it}^2\bigr\}$ is  symmetric  of dimension $T$, finite, positive definite, and diagonal. \label{item:het}
	\end{enumerate}
%with diagonal \mbox{elements $\sigma^2_{it}$} and zero elsewhere
	\end{enumerate}

%Ass. explanation
\par The above model assumptions have the following implications. \ref{item:dgp}.\ref{item:dgp_x} guarantees that the sequence of random variables $\{\xii' \xii \}$ is \emph{iid} [\textsc{prop 3.3} in \citet[p.30]{white1984}]. \ref{item:dgp}.\ref{item:dgp_u} imposes cross-sectional independence and, together with \ref{item:dgp}.\ref{item:dgp_x}, implies that $\{\xii' \errori \}$ is an \emph{inid} sequence of random vectors [\mbox{\textsc{prop 3.10}} in \citet[p.34]{white1984}]. Assumption \ref{item:dgp} and its implications remain unaltered after any data transformation.
%
%An implication of \ref{item:1way_error}  is $\error|\xii\distas{}NID(0,\sigma^2_{it})$, where $\sigma^2_{it}= \sigma^2_{\alpha}+\sigmaeps$.
%
\par The strict exogeneity assumption \ref{item:xu}.\ref{item:exogeneity} rules out feedback effects and implies contemporaneous exogeneity, i.e., $\E\bigl(\tildeuit|\Xii\bigr)= \E\bigl(\tildeuit|\xit \bigr)=0$, and is a crucial assumption to prove consistency of the \emph{within-group} estimator. The projection analog of \ref{item:xu}.\ref{item:exogeneity}  is the strong exogeneity condition, i.e, $\E\bigl(\widetilde{\mathbf{x}}_{is}\tildeuit)=0 \Leftrightarrow \E\bigl(\tildeuit|\Xii\bigr)\mbox{=0}$, for all $s\in \mathcal{T}$ and $s\ne t$. Because the exogeneity of the non-demeaned variables might not be strong enough to guarantee that exogeneity is preserved \emph{after} the transformation\footnote{This occurs because the regressor is correlated with $T\inv\iotat\iotat'\uii$ since it includes the whole history.}, i.e., $\E\bigl(\xii'\errori\bigr)\mbox{=0} \not\Rightarrow  \E\bigl(\Xii'\uii\bigr)\mbox{=0}$ \citep[p.707]{cameron2005}.  %The assumption is an implication of  \ref{item:1way_error} such that $\E(\error|\xii)=\E(\alpha_i|\xii)+\E(\epsit|\xii)=0$.
Assumption  \ref{item:xu}.\ref{item:het} allows the conditional error variance to vary across observations and time periods, and  imposes serial uncorrelation over time dimension, $\E\bigl(\tildeuit \widetilde{u}_{is}|\Xii\bigr)=0$ with $ (t,s)\in \mathcal{T}$ and $ t \ne s$.

\par The assumptions for the existence and optimality properties of the estimator of the true population parameter $\bm{\beta}$ are 

\begin{enumerate}[label=\sc{asm.}{\arabic*}, leftmargin=1.7\parindent, rightmargin=.5\parindent]
\setcounter{enumi}{2}
\item (\emph{rank condition}): $\Sxx \equiv N\inv \sum_{i=1}^N \Xii'\Xii$ is a finite symmetric matrix with full column rank $k$. \label{item:rank}

\item (\emph{moment conditions}):  \label{item:mom} %%
 	\begin{enumerate}[label={\roman*}]		
 	 	\item  $\E\big\|\Xii'\Xii\big\|<\infty$ for  $\Xii'\Xii\in \mathbb{R}^{k\times k}$;  \label{item:mom_x2} 			
 	 	\item  $\mathrm{sup}_{i}\,\E\big\|\Xii' \uii\big\|^{2+\delta}<\infty$ for some $\delta>0$, $\forall i$ and $\Xii' \uii \in \mathbb{R}^k$, \label{item:mom_Adistr}	%uniform mom bound
	\end{enumerate}	
 \noindent where $\|\cdot\|$ denotes the Euclidean norm.

\item  (\emph{average variance-covariance matrix convergence}): \label{item:V} \\
$\bVn=N\inv \sum_{i=1}^N \, \Vi\to\Vv$, where $\Vi=\E\bigl(\Xii'\Sgma\Xii\bigr)$ and $\Vv$ is a finite positive definite $k\times k$ matrix.
\end{enumerate}

%Ass. explanation
The full column rank condition in \ref{item:rank} implies non-singularity of the matrix $\Sxx$ and, hence, no perfect multicollinearity that guarantees the invertibility of the matrix.  The limiting matrix $\SSXX \equiv \E\bigl(\X'\X\bigr)$ possesses the properties of $\Sxx$ by the \emph{Weak Law of Large Numbers} (\emph{WLLN}) [\mbox{\textsc{thm 6.6}}].  Another implication of \ref{item:rank} is that the matrix of regressors $\Xii$ is full column rank.
\mbox{Assumption \ref{item:mom}} defines the finiteness and boundedness of moments in terms of the Euclidean norm. 
Assumption \ref{item:V} ensures that the average variance-covariance matrix converges to a finite quantity, satisfying one of the conditions of the \emph{Multivariate Central Limit Theorem} (\emph{MCLT}) for \emph{inid} processes  [\textsc{thm 6.16} in \citet[p.189]{hansen2019}].

No restrictions are placed on influential data points -- such as, high leverage points that possess extreme values in the covariates -- but we possibly allow for their presence. We consider a framework where the panel is small, that is, the time period length is smaller than the number of units $N$ such that $T \ll N$. Under this notation $T$ is the full set of time information, and the total number of observations in the sample is  given by $n = N\cdot T$ with balanced data sets. 

This set of assumptions and their implications remain valid under any monotonic data transformation due to the \emph{Continuous Mapping Theorem} (\emph{CMT}) [\textsc{thm 6.19} in \citet[p.192]{hansen2019}]. Later in this work, we consider the \emph{within-group} transformation of the data.

%%%%%%%%%%%%%%%%%%%%%%%
\subsection{Asymptotic Properties of the Estimator}\label{sec:properties}
 Under \ref{item:dgp}--\ref{item:mom}.\ref{item:mom_x2}, the \emph{within-group} estimator of the true population parameter $\bm{\beta}$ exists with form $\bfen= \Bigl(N\inv \sum_{i=1}^N \Xii'\Xii \Bigr)\inv \,\, N\inv\sum_{i=1}^N  \Xii'\yi$, and is consistent, i.e.,
\begin{equation}\label{eq:bfe_asymptotic}
\bfen -  \bm{\beta} = \Biggl(\frac{1}{N} \sum_{i=1}^N \Xii'\Xii \Biggr)\inv \,\, \frac{1}{N} \sum_{i=1}^N \Xii'\uii \overset{p}{\to}\bm{0} \,\,\, \mathrm{as} \,\,\,N\to \infty.
\end{equation}

%\begin{addmargin}[3em]{3em}
%Proof existence & consistency
%block A: X'X
\noindent The consistency of the \emph{within-group} estimator under the aforementioned assumptions is a known result \citep[as reference, see][pp.~612--613]{hansen2019}.
%%%
By the previously discussed implication of \ref{item:dgp}.\ref{item:dgp_x} and \textsc{prop 3.3} in \citet[p.30]{white1984}, $\bigl\{\Xii'\Xii\bigr\}$ is an \emph{iid} sequence of random variables with finite moments given \ref{item:mom}. The elements of the sequence satisfy the \emph{Weak Law of Large Numbers} (\emph{WLLN}) [\mbox{\textsc{thm 6.6}} in \citet[p.182]{hansen2019}] such that \mbox{$\Sxx \overset{p}{\to} \SSXX<\infty$}. Because both matrices are invertible by \ref{item:rank}, then \textsc{thm 6.19} [\emph{Continuous Mapping Theorem (CMT)} in \citet[p.192]{hansen2019}] yields the result \mbox{$\Sxx\inv \overset{p}{\to} \SSXX\inv$}. 
%block B: X'u

Now, we show that the second component in \eqref{eq:bfe_asymptotic} converges in probability to zero. We know that the sequence $\bigl\{\Xii'  \uii\bigr\}$ is \emph{inid} as an implication of \ref{item:dgp} [\textsc{prop 3.10} in \citet[p.33]{white1984}]. Then, the Chebyshev inequality is
\begin{equation}\label{eq:chebyshev}
\mathrm{Pr}\Biggl(\Bigg\| \frac{1}{N} \sum_{i=1}^N \Xii' \uii\Bigg\|\ge \epsilon\Biggr) \le \frac{\E\Big\|\frac{1}{N} \sum_{i=1}^N \Xii' \uii \Big\|^2}{\epsilon^2},
\end{equation}

%%%%%
\noindent where the numerator in \eqref{eq:chebyshev} can be expanded as follows
\begin{equation}\label{eq:induced_norm}
\begin{split}
\Bigg\|\frac{1}{N} \sum_{i=1}^N \Xii' \uii \Bigg\|^2 & = \tr\Biggl\{\Biggl(\frac{1}{N} \sum_{i=1}^N   \Xii'\uii\Biggr) \, \Biggl(\frac{1}{N} \sum_{j=1}^N \errorj'\xij\Biggr) \Biggr\}\\
& = \frac{1}{N^2}\, \tr\Biggl\{\sum_{i}\sum_{j}  \Xii'\uii\errorj'\xij\Biggr\}.
\end{split}
\end{equation}

By the aforementioned implication of \ref{item:dgp}.\ref{item:dgp_u} and under \ref{item:xu}.\ref{item:het} the conditional error variance is
\begin{equation}\label{eq:system}
\E\bigl(\Xii'\uii\errorj'\xij\big| \Xii \xij\bigr) = 
\begin{cases}
\zero & \forall i \ne j\\
\Xii'\Sgma\Xii &\forall i= j
\end{cases}
\end{equation}
%\par Set  the average variance-covariance matrix $\bVn=N\inv\sum_{i=1}^N\Vi$, where  $\Vi=\E\bigl(\xii'\Sgma\xii\bigr)$  is a finite, symmetric and positive definite matrix by construction.  The matrix $\bVn$ has the properties of $\Vi$. Let $\lambda_l(\bVn)$ be the $l\th$ eigenvalue of $\bVn$. 
\par Applying the expected value operator to \eqref{eq:induced_norm}, and using result \eqref{eq:system} jointly with the \emph{Law of Iterated Expectations} (LIE), the above equality becomes as follows
\begin{equation}\label{eq:exp_induced_norm}
\begin{split}
\E\Bigg\|\frac{1}{N} \sum_{i=1}^N \Xii' \uii \Bigg\|^2 & = \frac{1}{N^2}\, \tr\Biggl\{\sum_{i}\sum_{j} \, \E\bigl(\Xii'\uii\errorj'\xij\bigr)\Biggr\} \\
& = \frac{1}{N^2}\, \tr \Biggl\{\sum_i\sum_j \, \E\Bigl[\E\bigl(\Xii'\uii\errorj'\xij\big| \Xii,\Xjj\bigr)\Bigr]\Biggr\} \\
& = \frac{1}{N^2}\, \tr \Bigl\{\sum_i \, \E\Bigl[\Xii'\Sgma\Xii \Bigr]\Bigr\} \\
& = \frac{1}{N}\, \tr \biggl\{ \frac{1}{N}\sum_i \, \E\Bigl[\Xii'\Sgma\Xii \Bigr]\biggr\} \\
& = \frac{1}{N}  \tr \bigl\{\bVn\bigr\} \to 0, \hspace{2mm} \mathrm{as} \,\, N \to \infty
%& = \frac{1}{N} \sum_{l=1}^K \lambda_l(\bVn)\\
%& \to 0 \,\,\,\, \mathrm{as} \,\, N \to \infty,
\end{split}
\end{equation}
\noindent  since assumption \eqref{item:V} implies that $ \tr \bigl\{\bVn\bigr\} \to \tr \bigl\{\Vv\bigr\} $, which is finite. 
\par As a result, the right-hand side of Equality~\eqref{eq:exp_induced_norm} converges in probability to zero. So does the left-hand side. Inequality \eqref{eq:chebyshev} becomes

\begin{equation*}
\mathrm{Pr}\Biggl(\Bigg\| \frac{1}{N} \sum_{i=1}^N \Xii' \uii\Bigg\|\ge \epsilon\Biggr)\to 0 \,\,\,\, \mathrm{as} \,\, N \to \infty,
\end{equation*}
\noindent and, hence, $N\inv  \sum_i\Xii'\uii\overset{p}{\to} \bm{0}$.
%blocks A&B
\noindent By \textsc{thm 6.19} [CMT in \citet[p.192]{hansen2019}] , the result follows $\bfen - \bm{\beta} \overset{p}{\to} \SSXX\inv \cdot \bm{0} = \bm{0}$, or alternatively $\bfen \overset{p}{\to} \bm{\beta}$. This result holds for any monotonic transformation of the data. %Using \textsc{thm 6.19} again, we get the wanted result $\tbfen \overset{p}{\to} \bm{\beta}$.
%\end{addmargin}

Under  \ref{item:dgp}--\ref{item:V}, the estimator has the known asymptotic distribution below 
\begin{equation}\label{eqn:avar}
\sqrt{N}\bigl(\bfen - \bm{\beta}\bigr) \overset{d}{\to}\N \bigl(\bm{0},\SSXX\inv \Vv\SSXX\inv\bigr)\hspace{4mm} \mathrm{as} \,\, N \to \infty \text{ and } T \textrm{ fixed}.
%\sqrt{N}\bigl(\bfen - \bm{\beta}\bigr) = \Biggl(\frac{1}{N} \sum_{i=1}^N \Xii'\Xii \Biggr)\inv \,\, \frac{1}{\sqrt{N}}\sum_{i=1}^N  \Xii'\uii  \overset{d}{\to}\N\bigl(\bm{0}, \SSXX\inv\Vi\SSXX\inv\bigr).
\end{equation}
% Proof Avar
%\begin{addmargin}[3em]{3em}
\noindent A reference for this result is \citet[][pp.~624--625]{hansen2019}. The left-hand-side of Equation~\eqref{eqn:avar} can be re-written as follows
%%%%%%%%%%%%%%%%%%%%%
\begin{equation*}
\small
\sqrt{N}\bigl(\bfen - \bm{\beta}\bigr) = \Biggl(\frac{1}{N} \sum_{i=1}^N \Xii'\Xii \Biggr)\inv \,\, \frac{1}{\sqrt{N}}\sum_{i=1}^N  \Xii'\uii.
\end{equation*}
%%%%%%%%%%%%%%%%%%%%%

%block A: X'X
The sequence of random variables $\bigl\{\Xii'\Xii\bigr\}$ is \emph{iid} as implication of \ref{item:dgp}.\ref{item:dgp_x} and by \textsc{prop 3.3}  in \citet[p.30]{white1984}. With analogous arguments as those used above to prove consistency, $\Sxx\inv \overset{p}{\to} \SSXX\inv$. 
%block B: X'u
Under assumptions \ref{item:dgp} and \ref{item:xu}.\ref{item:exogeneity} and by \textsc{prop 3.10} in \citet[p.33]{white1984}, the sequence $\bigl\{\Xii'  \uii\bigr\} \in \R^k$ is \emph{inid} with means $\E\bigl(\Xii'\uii \bigr)=0$ and variance matrices \mbox{$\Vi=\E\bigl(\Xii'\Sgma\Xii\bigr)$}, by LIE and \ref{item:xu}.\ref{item:het}. The limit in probability \ref{item:V} and assumption \ref{item:mom}.\ref{item:mom_Adistr} are the two conditions that satisfy  the  \emph{Multivariate Central Limit Theorem} (\emph{MCLT}) for \emph{inid} processes  [\textsc{thm 6.16} in \citet[p.189]{hansen2019}]. Therefore, $N^{-1/2}\,\sum_{i=1}^N\Xii'\uii \overset{d}{\to}\N \bigl(\bm{0}, \Vv\bigr)$ as $N\to\infty$. 
%blocks A&B
\emph{Slutsky's Theorem} [\textsc{thm 6.22.2} in \citet[p.193]{hansen2019}] yields the result $\sqrt{N}\bigl(\bfen - \bm{\beta}\bigr) \overset{d}{\to} \N\bigl(\bm{0},\SSXX\inv \Vv\SSXX\inv\bigr)$, where $\SSXX\equiv\E\bigl(\Xii'\Xii\bigr)$. \textsc{thm 6.19} [\citet[p.192]{hansen2019}]  ensures that the above limits hold for any monotonic transformation of the data, e.g., the within-group transformation.

%%%%%%%%%%%%%%%%%%%%%%%%%%%%%%%%%%%%%%%%%%%%%%%%%%%%%%%%%%%%%%%%%%%%%%%%%%%%%
\section{Estimating the Asymptotic Variance}\label{sec:asyvar}
 %\red{Cattaneo 2018}
Given the above results under the model assumptions we made, the approximate distribution of the estimator of  $\bm{\beta}$ for large but finite samples is
\begin{equation}
\bfen\distas{a} \N(\bm{\beta}, N\inv\SSXX\inv\Vv \SSXX\inv),
 \end{equation}
 \noindent where the limiting matrices $\SSXX$ and  $\Vv$ need to be estimated, and  so does the average variance-covariance matrix $\bVn$.  While $\SSXX$ is estimated by $\Sxx = N\inv \sum_{i=1}^N \Xii'\Xii$, the estimation  of the average variance-covariance matrix needs further discussion. According to  \citet{white1980}, a computationally feasible practice consists in estimating each expectation, $\Vi=\E\bigl(\Xii'\Sgma\Xii\bigr)$, individually, and a plausible estimator of $\bVn$ would be $ N\inv \sum_{i=1}^N\Xii'\uii\uii'\Xii$ if the error term were known. Because it is unobserved, a consistent estimator of the  variance-covariance matrix is in practice $N\inv \sum_{i=1}^N\Xii'\tui\tui'\Xii$, where  \mbox{$\tui =\yi - \Xii\bfe$}.  Define $\tui = \resi$ to simplify the notation. 
 
Using a generalised expression for regression residuals, the  variance-covariance matrix can be re-written as follows: $\Vh= N\inv \sum_{i=1}^N\Xii'\vi\vi'\Xii$, where $\vi=\Mi\inv\tui$ are the transformed regression residuals with $\Mi$ being the transformation matrix that differs across estimators of the variance-covariance. When the transformed residuals equalise the residuals from the regression, $\vi=\I_T\resi$, the variance-covariance matrix  takes the familiar \emph{``sandwich-like''} formula of \mbox{\citeapos{arellano1987}} estimator. 
 
The  variance-covariance matrix $\Vh$ with transformed residuals is still a consistent estimator of the true variance.   Let $\Sgmahat=\vi\vi'$, from \citeapos{white1980} general result and under  the above model assumptions and \textsc{thm}~7.7 in \citet[p.232]{hansen2019}, it follows that \mbox{$\big\|\Vh-\bVn\big\|\overset{p}{\to}\bm{0}$}  and, hence,  $\big\|N\inv\sum_i\Sgmahat-\bSgma\big\|\overset{p}{\to}\bm{0}$, for all $i=1,\dots,N$, as $N\to\infty$ and keeping $T$ fixed. 

%The estimator of variance of form $\Vh$ mirrors the EHW estimator used for robust inference in cross-sectional models, and its first formularisation has been suggested by \citet{arellano1987} following \mbox{\citet[Ch.6]{white1984}}.  

%The EHW estimator and the other HC estimators of the variance were originally formalised for cross-sectional models. 

In the next sections, we review \citeapos{arellano1987} well-known formula, formalise  \citeapos{mackinnon1985}  jackknife-type estimator for panel data,  provide a panel version of \citeapos{davidson1993} estimator, and propose a new hybrid estimator, $PHC6$. The consistency of estimators with transformed residuals is derived in Appendix~\ref{app:trans_res}.

%%%%%%%% NEW SECTION
\subsection{HC$k$-type Estimators}\label{sec:estimators}
The well-known formula of \citeapos{arellano1987} estimator (henceforth, PHC0)
is  
\begin{equation}\label{eq:PHC0}
\avar_0=c_0\,\Sxx\inv\Vh^0\Sxx\inv\,,
\end{equation}
\noindent where $c_0=\frac{n-1}{n-k}\cdot\frac{N}{N-1}$, and $\Vh^0 = N\inv\sum_{i=1}^N \Xii' \vi \vi'\Xii$ with $\Mi=\I_T$. The finite-sample correction factor\footnote{Computationally, statistical software, like \textsc{stata}, use a finite-sample modification of the conventional (i.e., \citeapos{arellano1987}) variance-covariance matrix  multiplying $N\inv\sum_{i=1}^N \Xii' \resi \resi'\Xii$ by the correction factor \mbox{$c=\frac{n-1}{n-k}\cdot\frac{N}{N-1}$}, where $n=N\cdot T$  for one-way clustering in panel data, otherwise cluster-robust standard error turn out to be downward biased  \citep{arellano1987, bertrand2004, cameron2011}.},  $c_0$, ensures that $\Vh^0$ is consistent under \ref{item:xu}.\ref{item:het} with fixed $T$; the ratio $N/(N-1)$ is a  computational necessary degree-of-freedom correction to control for individual correlation \citep{stock2008,cameron2011}.  
\par  The estimator that resembles \citeapos{davidson1993} HC3 in the panel data framework (PHC3) is as follows
\begin{equation} \label{eq:PHC3}
\avar_3=c_3\,\Sxx\inv\Vh^3\Sxx\inv\,,
\end{equation}
\noindent where $c_3=(N-1) N\inv$, $\Vh^3 =\frac{1}{N}\sum_{i=1}^N \Xii'  \vi \vi'\Xii$ with $\Mi= (\I_T-\Hi)$ and the individual leverage matrix\footnote{The individual leverage matrix is a $T\times T$ matrix defined as follows %
\begin{equation}\label{eqn:Hi}
	\Hi = 
	\begin{pmatrix}
	h_{i11}&h_{i12}&\dots&h_{i1T}\\
	h_{i21}&h_{i22}&\dots&h_{i21T}\\
	\vdots&\vdots&\vdots&\vdots\\
	h_{iT1}&h_{iT2}&\dots&h_{iTT}	
	\end{pmatrix}
	\text{ for all } i = 1,\dots,N
\end{equation}
\noindent with elements $\hits = \xit'(\X'\X)\inv\xis$ with $t,s=1,\dots,T$.
}, 
\mbox{$\Hi = \Xii \bigl(\X'\X\bigr)\inv\Xii'$}, whose diagonal elements $\hitt = \xit'(\X'\X)\inv\xit$ lie in the $(0,1)$ interval but the off-diagonal elements may be negative. Predicted residuals, $\vi$, assign a penalty to LS residuals based on the degree of leverage making the estimates of the variance less sensitive to leverage points. This type of standard errors tend to be asymptotically conservative  as the number of covariates is allowed to grow as fast as the sample size, despite being asymptotically valid \citep{cattaneo2018}. 
%n = n in  Cattaneo(2018) p16

\par The estimator of the jackknife asymptotic variance for panel data models (PHC$jk$) adapts \citeapos{mackinnon1985} HC$jk$ estimator and has form 
\begin{align}\label{eq:jack}
%\Vh^{jk}  = \biggl(\frac{N -1}{N}\biggr) \biggl\{\Vh^3 -  \frac{1}{N}\X' \uh\uh'\X\biggr\} ,
\Vjk 	 %
&= \Biggl(\frac{N -1}{N}\biggl) \sum_{i=1}^N \Bigl( \bi - \bmean\Bigr) \Bigl(\bi - \bmean\Bigr)' \notag \\
&=\Biggl(\frac{N -1}{N^2}\biggl)\Sxx\inv   \Bigl\{\Vh^3 -  \mustar\mustar{'} \Bigl\}\Sxx\inv\,,
\end{align}
\noindent where the Leave-One-Out estimator is $\bi= \bfe- \bigl(\X'\X\bigr)\inv \Xii' \vi$ with $\Mi= (\I_T-\Hi)$, $\bmean =\frac{1}{N} \sum_{i=1}^N\bi$, and $\mustar=\frac{1}{N}\sum_{i=1}^N\Xii'\vi$. The jackknifed variance-covariance estimator with \emph{fixed effects} can be found in \citet{belotti2020}. 

In practice, the jackknife procedure consists in deleting the entire history of each unit one at a time without replacement. Because  the jackknife resamples  in such a way to construct ``pseudo-data'' on which the estimator of interest is tested, this technique -- as well as the bootstrap -- is suitable for the assessment of  the variability of an estimate, e.g., the estimation of standard errors \citep[Chapter~6]{efron1982,freedman1984}.  The advantages of the jackknife procedure are double: it is an entirely data-driven approach, and it is able to alleviate the impact of influential units on inference \citep{cattaneo2019}. The main drawback is that the jackknife estimator becomes computationally infeasible  for sufficiently large number of groups. 

PHC3 is a special case of Equation~\eqref{eq:jack} when the contribution of the second block is null as $N\to\infty$ and fixing $T$. The two estimators are asymptotically equivalent and coincide in sufficiently large samples. The derivation of \eqref{eq:jack} involves considerable algebraic manipulations  (see Appendix~\ref{app:l1o}).

%The jackknifing procedure is a data-driven method that can approximate the variance of the estimator $\bfe$ sufficiently well even with many covariates \citep{cattaneo2019}. 

%The development of the jackknife methodology traces back to \citet{efron1982} who introduced this approach  to correct for finite sample bias of an estimator. Because  the jackknife resamples  in such a way to construct ``pseudo-data'' on which the estimator of interest is tested, this technique -- as well as the bootstrap -- is suitable for the assessment of  the variability of an estimate -- i.e., for the estimation of standard errors\footnote{In practice, applications with jackknife standard errors have not found as fertile grounds as the bootstrapped standard errors.} --  with only the data at hand \citep[Chapter~6]{efron1982,freedman1984}.  In detail, the procedure consists in deleting a single or a group of observations one at a time without replacement. In the panel framework, the jackknife estimator is constructed in a way to delete the whole history of each unit\footnote{This is similar to deleting $g$-groups of observations in cross sectional studies.} one at a time. The advantages of the jackknife procedure are double: it is an entirely data-driven approach, and it is able to alleviate the impact of influential units on inference \citep{cattaneo2019}.
%Belotti, Peracchi -- stata tool -- KEEP AN EYE ON IT !!!!!!!!!!!!!

\subsection{A Hybrid Estimator: PHC6}\label{sec:phc6}
\par We propose a hybrid estimator of the variance, PHC6, that nests PHC0 and PHC3 estimators using a threshold criterion from the decision rule of the penalty factor in \citet{cribari2004}. PHC3 is chosen because Monte Carlo simulations showed that  \citeapos{davidson1993} HC3 possess the best final sample properties in terms of lower bias, with  rejection rates closer to the nominal one \citep{long2000}. The threshold criterion is designed to account for the time period in which each unit has exerted the maximal leverage with respect to the average leverage in the same period.  PHC6 is designed to deliver standard errors that are  higher in magnitude than PHC0 with contaminated observations but the same as PHC0 standard errors with no extreme observations in the sample.  

%\footnote{\citeapos{cribari2004} HC4 estimator for cross-sectional models is \mbox{$\avar_4= \, (\XX'\XX)\inv\Bigl\{\frac{1}{N}\sum_{i=1}^N \frac{\hat{u}_i^2}{(1-h_{ii})^{\delta_i}}\textbf{x}_i\textbf{x}_i'\Bigr\}(\XX'\XX)\inv$}, where $\textbf{x}_i$ is $k\times1$, $\XX$ is $N\times k$, and $\delta_i = min\bigl\{4,Nh_{ii}/k\bigr\}$. This formula has not yet been extended to panel data models.} 

Before presenting the proposed estimator, we clarify beforehand the notation we will be using. Let the  $T\times1$ vector
\begin{equation*}
	\hi = diag(\mathbf{H}_{i}) = %
	\begin{pmatrix}
		h_{i11}\\
		h_{i22}\\
		\vdots\\
		h_{iTT}	
	\end{pmatrix}
	\text{ for all } i = 1,\dots,N
\end{equation*}
\noindent be the  individual leverage vector  constructed from the diagonal elements of the individual leverage matrix $\Hi$ defined in \eqref{eqn:Hi}, and let the $T\times1$ vector
\begin{equation*}
	\bHtt =
	\begin{pmatrix}	
		 \overline{h}_{11} \\
		\overline{h}_{22}\\ 
		 \vdots \\
		 \overline{h}_{TT}
	\end{pmatrix}
	= %
	\begin{pmatrix}
		N\inv\sum_{i=i}^N h_{i11}\\
		N\inv\sum_{i=i}^N h_{i22}\\
		\vdots\\
		N\inv\sum_{i=i}^N h_{iTT}	
	\end{pmatrix}
\end{equation*}
\noindent be constructed from  the average leverage at time $t$ across units. Then, let $\bhvec $ be a $T\times 1$ vector with elements $\big(\bHtt\exp\circ\bfj\big)$, where the expression $\exp\circ \bfj$ indicates the element-wise power of $\bfj$ which is a $T\times1$ vector  of negative ones. The Hadamard (element-wise) product,  $\hi \odot \bhvec$,  is a $T\times 1$ vector whose elements, $\hitt\bhtt\inv$, inform on the relative leverage of unit $i$ at time $t$ with respect to the average leverage at \mbox{time $t$}. Specifically, values of $\hitt\bhtt\inv$ above one signal that the relative leverage of unit $i$ at \mbox{time $t$} exceeds the average influence at \mbox{time $t$}. Units with values slightly grater than one cannot automatically be flagged as highly influential because in the absence of influential units at time $t$, the denominator may be very close to the numerator, by construction and, hence, one cannot be chosen as cut-off value. Conversely, high values of $\hitt\bhtt\inv$ indicate that unit $i$ is exerting high leverage at time $t$ with respect to the mean influence at time $t$. 

\par The PHC6 estimator of the variance is defined as follows
\begin{equation}\label{eq:PHC6}
\avar_6=c_6 \, \Sxx\inv\Vh^6\Sxx\inv\,,
\end{equation}
\noindent where the variance-covariance matrix is $\Vh^6 = \frac{1}{N}\sum_{i=1}^N \Xii' \vi\vi'\Xii,$ and the matrix $\Mi$ has functional form 
\begin{equation}\label{eq:Mi}
\Mi =%
\begin{cases}
\I_T  & \text{if }\histar<2\\
\I_T-\Hi & \text{otherwise} 
\end{cases}
\end{equation}
\noindent  where $\histar = max\big\{h_{i11}/\overline{h}_{11},\dots,h_{iTT}/\overline{h}_{TT}\big\}$ is the maximal individual leverage of unit $i$; and $\bhtt = N\inv\sum_{i=i}^N\hitt$ is the average leverage at time $t$, with $\hitt$ being the individual leverage of unit $i$ at time $t$. 
 The finite sample correction of PHC6 is
	\begin{equation*}
		 c_6=
		\begin{cases}
		\frac{(NT-1)N}{(NT-k)(N-1)}& \text{if } \histar<2\\
		\frac{N-1}{N}&  \text{otherwise} 
		\end{cases}
\\.
	\end{equation*}
 
%and $\mathrm{tr}({\Hi})$ is the trace of the projection matrix $\Hi$, that cannot be equal to the total number of regressors in the model, by construction. The  a small-sample correction factor $c_6$ is equal to $c_0$ when $\frac{n}{k}\cdot\frac{\mathrm{tr}({\Hi})}{T}<4$, and equal to $c_3 = (N-1) N\inv$ otherwise. 
%\par  The ratio $\mathrm{tr}({\Hi})/T$ is an average measure of leverage lying in the (0,1) interval. The cutoff value of 4 is the maximal penalty that \citeapos{cribari2004} HC4 estimator assignes to predicted residuals  in the presence of highly influential observations. %The matrix-vector product $\Mi\resi$ is a $T\times1$ vector of predicted residuals.

 According to the cut-off rule,  residuals of units with maximal individual  relative leverage, $\histar=\hitt\bhtt\inv$, are discounted by the penalty matrix $\Mi$.  Unlike PHC3 that penalises  both low and high leverage points at the same rate, PHC6 discounts at the same discounting rate as PHC3 only if the unit exerts high leverage. When the individual  relative  leverage does not exceed the cutoff, no penalty is applied and PHC6 coincides with \citeapos{arellano1987} estimator. Conversely, when the average level of leverage exceeds the cut-off value, PHC6 residuals are penalised as in PHC3.  In addition, PHC6 always weights for a final sample correction. 
\par  The cut-off is set to be equal to 2 such that no penalty is assigned to fairly influential units at time $t$. One is not chosen as a cutoff value because  in the absence of anomalous cases, the denominator $\bhtt$ would be very close to the numerator $\hitt$ for some units with meaningless individual leverage but above the mean average. This would drive the ratio to exceed one.

\section{Monte Carlo Simulation}\label{sec:mc_sim}
%%%%
%%[OLD VERSION W/2SIMUL] 
%\par We design two types of Monte Carlo simulations to illustrate how the aforementioned estimators of the variance behave in finite samples using two sets of simulated data to resemble the data complexity and variety. In the following sections, we  present two alternative simulation designs and the relative results in terms of performance (i.e., size of test and power of test) of HC estimators. Specifically,  one simulation set up uses synthetic balanced data and does not allow for any correlation between the individual-specific fixed effects and the regressor\footnote{This design leaves open the possibility to estimate the regression equation consistently and efficiently using the \emph{random effects} (RE) estimator.} while the other makes use of unbalanced simulated data\footnote{The unbalanced data sets are generated by randomly dropping observations from the sample as explained in Section~\ref{sec:simuldesign}. The discussion of obsessions that are missing at non-random is postponed to future analysis when we will be focusing on attrition in panel data.} and assumes correlation between the unobserved heterogeneity (i.e., the fixed effect term) and the first regressor.

In this section, we  present the MC simulation design which illustrates the behaviour of the four types of estimators of the variance in finite samples\footnote{Monte Carlo simulations provide computational evidence of finite sample properties of an estimator or a test when applied to fictitious data \citep{hendry1984, kiviet2012}.}, when variables are contaminated with anomalous data points.   %Because simulation results highly depend on the assumptions made and on the calibrated parameters of interest, it is the role of the researcher to design simulations that resemble real data complexity and problematics as precisely as possible. % \citep[ch.6]{hendry1984}
For simplicity,  the simulation set up uses synthetic balanced data set and does not allow for any correlation between the individual-specific fixed effects and the regressor\footnote{This design leaves open the possibility to estimate the regression equation consistently and efficiently using the \emph{random effects} (RE) estimator. However, our objective is not to analyse RE because its assumptions are unlikely to be satisfied in practice. Also, we are not focusing on unbalanced datasets, whose discussion is postponed to future analysis while addressing the issue of attrition in panel data.}.
The data generating process for the Monte Carlo simulation is designed to be closely related to: (i) \citet{godfrey2006}, \citet{stock2008}, and \citet{mackinnon2013} in terms of the form of heteroskedasticity, number of regressors and the calibrated parameters; and (ii) \citet{bramati2007} for the contamination with cell-isolated good leverage points. However, we depart from these settings by making some modifications to the simulation designs. %High leverage units are randomly generated as explained in Section~\ref{sec:simuldesign}.

%mc
The data generating process (DGP) of Monte Carlo simulations is as follows
%for is based on a balanced simulated panel dataset generated as follows
\begin{align}
& \y = \beta_0+  \sum_{k=1}^K \beta_j x_{it,k}+ \alpha_i+ \error,  \text{ for all } i=1,\dots,N \text{ and } t=1,\dots,T_i \label{eq:specification} \\ 
%%%%
%& \text{ either } \alpha_i \distas{iid} \mathrm{U}(0,1) \text{ or } \alpha_i = \frac{1}{T_i}\sum_{t=0}^{T_i} x_{it,1}  \hspace{5pt} \\
%&\x\distas{} \N(\bmu ,\bm{\Sigma}_x),  \,\, \textrm{with}  \,\, \bmu=(\mu_1,\mu_2)'   \,\, \textrm{and}  \,\,\bm{\Sigma}_x = \diag\{\sigma_1,\sigma_2\}\, \textrm{for} \, k = 1,2 \text{\footnotemark} \\
&x_k\distas{} \N(0,1)   \,\, \textrm{for $k=\{1,2\}$ except contaminated cases}  \\
&x_k = f(x_1,x_2) \,\, \textrm{for} \,\, k = \{3,4,5\}\\
&\alpha_i \distas{}{}U(0,1) \label{eqn:alpha}\\ 
%%%%
&   \error= \sigma_{it} \epsilon_{it} + \theta \epsilon_{i t-1}, \,\, \epsilon_{it} \distas{} \N \bigl(0,1\bigr), \,\, \error \distas{} \N \bigl(0,\sigma_{it}^2\bigr)  \\
%%%
&\sigma_{it}^2 = z(\gamma) \Biggl(\beta_0+ \sum_{j=1}^J \beta_j x_{it,j} \Biggr)^{\gamma}, \hspace{5pt} \mathrm{with} \,\, z(\gamma)= \Biggl[\E \Biggl(\beta_0 + \sum_{j=1}^J \beta_j x_{it,j}\Biggr)^{\gamma}\Biggr]\inv
%%%
%& \x\distas{iid} \N(0,1), \hspace{5pt} \text{except  the  contaminated  cases.} 
\end{align}
%\footnotetext{The contamination is completely random over the observations and is explained below.}
\noindent where the number of regressors in the model is $K=5$ and $K=J$;  model parameters are calibrated to be $\beta_{k}=1$, \mbox{for $k=1,\dots,4$}, and  $\beta_5=0$; $\theta=0$ because errors are conditionally serially uncorrelated by assumption as in \citet{stock2008}; the degree of heteroskedasticity assumes values of $\gamma = \{0,2\}$, where $\gamma=0$ stands for homoskedasticity and  $\gamma\gg1$ for severe heteroskedasticity. The scaling factor, $z(\gamma)$, is chosen such that the average variance of  the error term is equal to one\footnote{The error term $\error$ is intrinsically heteroskedastic but not on average due to the presence of the scaling factor $z(\gamma)$. The distribution of  the random variable $\W=\beta_0+\sum_{j=1}^J \beta_j x_{it,j}$ and $\W^2$  is provided in Appendix~\ref{sec:w}. The algebraic derivation of the means and variances are shown.}. 
 %\textrm{\footnote{\textrm{Their values vary for uncontaminated and contaminated cases. Below we explain how we contaminated the data.}}}
\par The contamination of random variables with good leverage points is completely random over the observations (i.e., \emph{cell-isolated} anomalous cases). Good leverage points are obtained by randomly replacing 10\% of the values\footnote{The degree of contamination could have been set to be even more or less severe according to the relevance the researcher attributes to  the presence of extreme observations in the sample. } of $x_1$ with extreme observations drawn from a normal distribution with mean $\mu_{x_1}=5$ and standard deviation $\sigma_{x_1}=25$. Because $x_1$ is contaminated, then the variables generated from the former are directly affected by this source of contamination.  The remaining random variables -- $x_3, x_4, x_5$ -- are either generated from the square or the product of $x_1$ and $x_2$ and, hence,  follow a $\chi^2_{(\nu_1)}$ and a Gamma distribution, respectively.

%%FE
\par The model is estimated including the set of aforementioned time-varying covariates and individual specific fixed effects, $\alpha_i$. We estimate model~\eqref{eq:specification} using fixed effects (FE) by applying  the \emph{within-group} (or time-demeaning) transformation to simulated data. Then,  we estimate the time-demeaned regression specification using OLS\footnote{%fgls
We do not  use the FGLS-FE to estimate the estimating Equation~\eqref{eq:specification} for three main reasons. First,  when the sample size is not sufficiently large there is an efficiency loss with respect to the FE-OLS estimator. In this analysis, we are interested in investigating the finite sample properties of  the estimator, when $N$ is not very large. Second, the FGLS-FE procedure requires to drop one of the time periods because the variance matrix is not invertible, leading to the reduction of the (already small) panel sample size \citep[ch.21.6, p.729]{cameron2005}. Third, FGLS-FE relies on the quality of the estimation of the variance and on the knowledge of the form of heteroskedasticity. However, the form of heteroskedasticity is always unknown from the data and the researcher has to make assumptions on the relationship between the variance of the disturbances and observables and unknown parameters \citep[Chapter~21, pp. 720-721, 729]{cameron2005}. This is unpractical in many areas of application and subjective to the researcher's guess. To overcome this limit, an objective criterion that has become a standard practice  in applied works consists in using conventional robust standard errors due to software facilities \citep{verbeek2008}.}.
As in \citet{hansen2007}, the DGP for the simulations involves only random effects (RE) model because  with~\eqref{eqn:alpha} we assumed that the unobserved fixed effect is uncorrelated with the regressors. The model could be estimated more efficiently with RE but FE models are commonly used in empirical studies with panel data\footnote{In future analysis we will re-assess the current version of the Monte Carlo simulation allowing $\alpha_i$ to be correlated with $\mathbf{x}_i$ to satisfy FE assumptions. Under this simulation design, $\bfe$ estimated with FE remains consistent  but is less efficient than $\bfe$ estimated with RE.}.
%Since this assumption is not often satisfied in practice, the analysis is conducted with FE-OLS. 

Our Monte Carlo simulation involves 10,000 replications.  The simulations are run for a combination of cross-sectional units $N=\{25, 50, 150, 500\}$ and time periods \mbox{$T=\{2, 5, 10, 20\}$.} Both cross-sectional units and time periods can be grouped as small ($N=\{25, 50\}$; $T=\{2,5\}$), moderately small ($N=\{150\}$; $T=\{10\}$), and moderately large ($N=\{500\}$; $T=\{20\}$). %We examine the cases of both severe heteroskedasticity, $\gamma=2$, and homoskedasticity, $\gamma=0$.  %Robust version of the variance-covariance  matrix is valid in the presence of heteroskedasticity provided that $T<N$, condition that is always satisfied in our simulation as our theoretical framework is based on the conventional finite-$T$ asymptotics \citep{stock2008,cattaneo2018}.
The simulation is programmed in \textsc{Stata}16-\textsc{mp}  and the main procedure is implemented in \textsc{mata}. %Cluster-robust and jackknife standard errors are programmed to correspond to \textsc{stata}'s  formulae.

%%%%%%%%%%%%%%%%%%%%%%%%%%%%%%%%%%%%%%%%
\section{Testing the Performance of HC Estimators}\label{sec:simulres}
\par  We examine the performance of each estimator in terms of  proportional bias (PB), rejection probability (RP, or empirical size), adjusted power test, and root mean squared error (RMSE). Results are provided for a battery of estimators by a combination of panel units, time periods, and degree of heteroskedasticity, $\{N,T,\gamma\}$, where the number of units $N$ varies in an interval from 25 to 500 units, time is fixed at $T=\{2, 5, 10, 20\}$, and the parameter that controls for the degree of heteroskedasticity is $\gamma\in\{0,2\}$. This design  is in accordance with the finite $T$ assumption in the model as time periods are fixed while the number of observations increases.

%We find that the downward bias of conventional cluster-robust standard errors does not vanish with the increase in the sample size, as documented in the cross-sectional literature, but it gradually reduces and its empirical size slowly converges  to the true $\alpha\%$-level. 
Good leveraged data and heteroskedasticity make, as expected,  test statistics calculated with conventional robust standard errors over-sized, upward biased, and with low power when the panel size $n<2,500$. The  proposed PHC6 mimics the behaviour of PHCjk in terms of PB, RP and power in all samples. PHC3 shows similar patterns but with different magnitudes.

\subsection{Rejection Probability and Probability Bias}\label{sec:test}
RP (i.e., the size of a test) in a Monte Carlo exercise with $R$ runs is the frequency at which a rejection of the true null hypothesis occurs on average. A test statistics has a good size if rejects the null hypothesis approximately around the chosen $\alpha\%$  of the simulations, when the model is generated under the assumption that the null hypothesis is actually true.  

%%%%%%%%%%%%
% 2sided - single test %
%%%%%%%%%%%%
\par The steps to obtain the empirical size in a two-sided single coefficient test are as follows. First, for each combination of $\{N,T\}$ and each simulation run $r = 1,\dots,R$,  compute the test statistics under the true null hypothesis, 
%%%
\begin{equation*}
T_{N,T}^0(\bfe_{N,T,r})= \frac{\bigl(\bfe_{N,T,r} - \bbeta^0\bigr)}{\sqrt{\AVar}}\distas{a} t_{(df_r, \,\alpha/2)}.
\end{equation*}
%%%
\noindent Second, set the indicator $\one\{\cdot\}$ to turn on when the null hypothesis is rejected according to the rule 
\begin{equation*}
J_{N,T,r}^0(\bfe) \equiv \one\bigl\{\big| T_{N,T}^0(\bfe_{N,T,r})\big| > t_{(df_r, \,\alpha/2)}\bigr\}, 
\end{equation*}
\noindent where $ t_{(df_r, \,\alpha/2)}$ is the critical value from a student-t distribution with $ df_r,$ degrees of freedom for a two-sided hypothesis test\footnote{With non-clustered inference $df_r=(NT-1)-(N+k-1)$ otherwise $df_r=N-1$.}. %In Stata this information is stored after the estimation in \texttt{e(df\_r)}
%%%
Third, count the  total number of times a rejection has occurred and average it out by the number of replications $R$; the empirical size denotes the percentage of rejections in the Monte Carlo exercise as
\begin{equation*}
\bar{J}_{N,T,r}^0(\bfe) \equiv\frac{1}{R} \sum_{r=1}^R J_{N,T,r}^0(\bfe) = \alpha_{test}.
\end{equation*}

%%%%%%%%%%%%
% 2sided - joint test %
%%%%%%%%%%%%
\noindent  For a two-sided test with $q$ linear restrictions, the coverage probability is computed as follows. First, for each combination of $\{N,T\}$ and each simulation run $r = 1,\dots,R$,  compute the Wald statistics under the true null hypothesis, $H_0: \mathbf{R}\bm{\beta}-\mathbf{r}^0=\bm{0}$, 
\begin{equation*}
W_{N,T,r}^0(\bfe) = N(\mathbf{R}\bfe_{N,T,r}-\mathbf{r}^0)' \Bigl\{ \mathbf{R}\AVar \mathbf{R}'\Bigr\}\inv ( \mathbf{R}\bfe_{N,T,r}-\mathbf{r}^0) \distas{a}\chi^2(q),
\end{equation*}
\noindent where  $\mathbf{R}$ is a $q\times K$ matrix with $q \le K$, and $\mathbf{r}^0$ is a $q\times1$ vector.
%%%%%%%%%%
\noindent Second, Mark as one every time a rejection occurs according to the rule 
\begin{equation*}
\tilde{J}_{N,T,r}^0(\bfe) \equiv \one\bigl\{W_{N,T,r}^0(\bfe) > cv_{\chi^2(q)}\bigr\}, 
\end{equation*}
\noindent where $cv_{\chi^2(q)}$ is the critical value from a $\chi^2$ distribution with $q$ degrees of freedom for a two-sided hypothesis test\footnote{Alternatively, the F statistic can be computed from the Wald test statistics as 
\mbox{$F_{N,T,r}^0(\bfe) = W_{N,T,r}^0(\bfe)/q\distas{a}F_{\alpha}(q, df_r)$} under the true null hypothesis, where $q$ are the number of restrictions and degrees of freedom at the numerator, and $df_r$ are the residual degrees of freedom or degrees of freedom at the denominator.}. %In Stata this information is stored after the estimation in \texttt{e(df\_r)}
%%%%%%%%%%%
\noindent Third, sum the cases when the null hypothesis has been rejected according to the above rule, and divide the number by the total number of simulation runs. The empirical size for a joint coefficient test is given by the percentage of rejections in the overall Monte Carlo as follows
\begin{equation*}
\bar{\tilde{J}}^0_{N,T,r}(\bfe) \equiv R\inv \sum_{r=1}^R \tilde{J}^0_{N,T,r}(\bfe) = \alpha_{test}.
\end{equation*}

%%%%%%%%%%%%%%%%%%%%%%%%%%%%%%%%%%%%%%%%%%%%
%% previous version %%
%%%%%%%%%%%%%%%%%%%%%%%%%%%%%%%%%%%%%%%%%%%%
%\par For a two-sided test of $q$ restrictions, the coverage probability is computed as
%\begin{equation}\label{eq:test} 
%%\frac{1}{R} \sum_{r=1}^R \large\one\Bigl(F^0 >F_{\alpha}(q, N-K)\Bigr)= \alpha_{test} \hspace{1.5mm} \textrm{under} \hspace{1mm} H_0, 
%Prob\Bigl(F^0 >F_{\alpha}(q, df_r)\Bigr)= \alpha_{test} \hspace{1.5mm} \textrm{under} \hspace{1mm} H_0, 
%\end{equation}
%%is the Wald statistics divided by the number of restrictions,
%\noindent where the statistic $F^0 = W/q\distas{a}F_{\alpha}(q, df_r)$ under the true null hypothesis; $q$ are the number of restrictions; $df_r$ are the residual degrees of freedom;
%and the Wald statistics is defined as $W = N(\mathbf{R}\bfe-\mathbf{r})' \Bigl\{ \mathbf{R}\widehat{\mathrm{AVar}(\bfe)} \mathbf{R}'\Bigr\}\inv ( \mathbf{R}\bfe-\mathbf{r}) \distas{a}\chi^2(q)$ under  $H_0: \mathbf{R}\bm{\beta}-\mathbf{r}=\bm{0}$, where  $\mathbf{R}$ is a $q\times K$ matrix with $q \le K$, and $\mathbf{r}$ is a $q\times1$ vector; and $F_{\alpha}(q, N-K)$ is the critical value from a \mbox{$F$ distribution} with $q$ degrees of freedom at the numerator and $ N-K$ at the denominator. 
%%%%%%%%%%%%%%%%%%%%%%%%%%%%%%%%%%%%%%%%%%%%
\par  In the simulations, we test $H_0: \beta_j=1$ against $H_1: \beta_j\ne1$ for $j=1$ while in a two-sided joint test we test $H_0: \beta_1=\beta_2=\beta_3=\beta_4=1$  against $H_1: \text{at least one } \beta_j\ne1$,  for $j=1,\dots,4$. The closer the rejection probability is to the nominal level of $5\%$, the better the estimator's performance in terms of empirical size (or type I error).
%%
%		qui gen pb`v'_`s' = 1 - (se`v'_`s'/std_b`v')		//PB>0 means that SE(\hat\beta) underestimates true SE

The proportional bias (PB) is a measure of the bias of the estimator of the variance-covariance matrix computed as $PB = 1 - SE(\hat{\beta}_j)/SD(\hat{\beta}_j)$, where \emph{SE} stands for standard error and \emph{SD} for standard deviation. Positive (negative) values of PB indicate by how much the standard error obtained using one of the four formulae presented above underestimates (overestimates) the ``true'' standard error. 

In this section, we comment on the performance of each estimator taking into account its ability to reject the true null hypothesis at 5\% significance level along with its accuracy. %We leave the assessment in terms of the RMSE for later discussion in Section~\ref{sec:rmse}
Tables~\ref{tab:simul1_v1} and~\ref{tab:simul1_v1homo} report the results of the Monte Carlo simulations respectively, with and without heteroskedasticity. Each table compares the PB, RP and RMSE\footnote{Results for the RMSE are commented in Section~\ref{sec:rmse}.} of four alternative  formulae of the variance-covariance matrix (i.e., PHC0, PHC3, PHC6 and PHCjk).  Results are  grouped by different combinations of sample size $N$ and time length $T$. Figures refer to the slope parameter $\beta_1$, which is associated with the contaminated variable $x_{it,1}$. The t-test statistics are at 5\%-level. 

%%%%%%%%%%%%%%%%%%%
%\input{ch1/mc_simul2_old}
%%HETEROSKEDASTICITY
%%PB
Under heteroskedasticity, PHC0 standard errors considerably underestimate the ``true'' variance (positive PB) on average by at least $30\%$ when $n\le2,500$. PHC6 mimics the behaviour of PHCjk in small and large samples, overestimating the true variance (negative PB: min= 1.2\% and max = 12.3\%) for $n\le300$ and slightly underestimating  the true variance (positive PB: min= 4.9\% and max = 10.6\%) in the other cases. For $N=\{25,50\}$ and all $T$ the PB of PHCjk is larger in absolute value than PB of PHC6 if the bias is positive, and smaller otherwise. From $N\ge150$ PHCjk and PHC6 produce the same bias but PHC3 produces a smaller bias in absolute value when the estimators over-estimate the variance.
%When PHC6 standard errors are upward biased in small samples (the true null hypothesis is not over-rejected  as often as PHC3. However, as the sample size increases PHC6 tends to mimic the behaviour of PHCjk, in terms of probability bias. Standad errors obtained with PHC3 and PHCjk formulae are much less biased than the other two alternatives, although their size is slightly distorted downwards but still the closest to the nominal $5\%$ level when $n<2500$.

%%RP
Test statistics of PHC0 are largely over-sized (RP above 0.05) when  $N=\{25,50\}$ and all $T$ but   approach the true $\alpha\%$-size when  $n\ge 5,000$, despite the high positive PB. The most conservative estimators always under-reject the null hypothesis (RP below 0.05), and as the cross-sectional size increases (fixing the time dimension) the RP gradually converges to 5\% but their test statistics still remain slightly under-sized. However, looking at the (positive/negative) distance from 0.05 PHC0 turns out to be more over-sized then he other estimators when $n\le750$.

In general, a smaller PB in absolute value (signaling a good approximation of the ``true'' variance) does not automatically imply that the empirical size is the closest to the actual nominal significance level. The ``true'' standard errors remain under-estimated (over-estimated) if the bias is positive, despite producing test statistics that reject the null hypothesis with much precision.
%It also produces test statistics that are, as expected, upward distorted.

%het b2
%\par \textbf{Heteroskedasticity and uncontamination.} PHC0 standard errors outperform all other estimators of the variance even though they display a greater (positive) bias but the empirical size of their test statistics is the closest to the nomial level, especially compared to those compited using PHC3 and PHCjk standard errors whose rejection probabilities touch 0\% with $N=\{25,50\}$ and for all $T$. On the other hand, PHC6 estimator performs almost as well as the conventional robust formula with $T=20$ and for all $N$. As the panl sample size increases, the rejection probabilities of PHC3, PHC6 and PHCjk standard errors tend to converge to the same value and, despite their smaller proportional bias, PHC0 test statistics have the most accurate empirical size.
%homo b1
%Under heteroskedasticity, results differ across simulation designs for the contaminated variable.%In both designs, similar results are obtained for the uncontaminated variable under heteroskedasticity as shown in  Tables~\ref{tab:simul1_v2} and \ref{tab:simul2_v2}. 

%With unbalaced panel data and correlation between the regressor and the unobserved heterogeneity, results for simulation of type 2 in Table~\ref{tab:simul2_v1} are in strack contrast with respect to those obtained for simulation of type 1 as

\par  Under homoskedasticity, PHC0 always underestimates the true variance (especially for $n\le1,500$). The PB reduces as the panel size increases but only when $n=10,000$ it drops considerably. The other PHC estimators tend to over-estimate the true variance (negative PB) but the magnitudes are smaller in absolute value than the figures of PHC0. PHC6 has similar bias to PHCjk while PHC3 is slightly more biased.
Test-statistics of PHC0 are over-sized (large RP) for $n\le1,500$ but show a convergence pattern to 5\% as the sample size increases. The test size of PHC6 and PHCjk is always closest to the true $\alpha$-size followed by PHC3. 

% similar conclusions can be drawn. The conservative estimators PHC3 and PHCjk have better performances in terms of RP and PB than PHC0 standard errors that understimate the true variance and over-reject the null hypothesis in almost all combinations of $\{N,T\}$. \red{PHC6 performs much better than the first two when $N\ge50$ and for all time periods.} PHC6 has the smallest bias in absolute value or the closest to PHCjk. The empirical size of PHC6 test statistics tends to becloser to the $\alpha\%$-value. As the sample size increases above 2500 units (for different combinations of  $\{N,T\}$), PHC0 behaviour shows some improvements.

%homo b2
%\par \textbf{Homoskedasticity and uncontamination.}  From Table~\ref{tab:simul1_v2homo}, PHC0 and PHC6 outperform the other two estimators in terms of rejection probabilities and, among the two, PHC6 is less upward biased with empirical size closer to $5\%$. Despite the smaller bias of PHC3 and PHCjk standard errors, their size is distorted downwards but as the smaple size increases, their size approxiamtes the nomial significance level with a smaller probability bias.

%joint
\par Tables~\ref{tab:wsimul_het} and~\ref{tab:wsimul_homo}  report the Wald test statistics and  RP from the joint coefficient test for the slope coefficients different from zero (i.e., $\beta_i$ for $i=1,\dots,4$)  under heteroskedasticity and homoskedasticity, respectively. Results for different combinations of $\{N,T\}$ are displayed. The nominal level of significance is set at $\alpha=0.05$. The closer the value of the rejection rate of the test statistic is to $\alpha=0.05$, the better the estimator's performance in terms of empirical size. 

Under heteroskedasticity and good leveraged data points, the RPs of the four estimators slowly converge to 5\% as the sample size increases with exception of PHC6.  PHC6 is outperformed by PHC0 in terms of RP for  $n\ge2,500$. Despite the upward distortion of all test statistics for $N\le500$ and all $T$, PHC0 raw sizes are the largest in magnitude among the four estimators in very small cross-sectional samples. The Wald statistics of the other two conservative estimators are the lowest in magnitudes for $n\ge300$. Similar patterns are observed under the assumption of homoskedasticity. PHC0 performs as well as the two conservative estimators only for large $N$ ($N\ge150$ fixing $T$).

\subsection{RMSE Assessment}\label{sec:rmse}
An additional evaluation on the quality of the four estimators is done in terms of the RMSE. For each estimator of the variance, the  RMSE is computed as the square root of the average deviation of the standard error from the standard deviation of the estimated coefficient of  $\beta_j$. In formulae, 
\begin{equation}
\mathrm{RMSE}_j^s=\frac{1}{R}\sum_{r=1}^{R}\sqrt{(\hat{\sigma}^s(\hat{\beta_j})_r-\sigma(\hat{\beta_j})_r)^2}
\end{equation}
\noindent where  $\hat{\sigma}^s(\hat{\beta_j})_r$ is the standard error of $\hat{\beta_j}$ in the $r$th run of the simulation computed using one of the HC formulae, and $\sigma(\hat{\beta_j})_r$ is  the standard deviation of the estimated coefficient $\beta_j$. A good quality estimator has its RMSE close to zero. Because the  RMSE and PB are  constructed from the same quantities, $\hat{\sigma}^s(\hat{\beta_j})$ and $\sigma(\hat{\beta_j})$, they are linked one to the other.  The larger the proportional bias in absolute value, the larger the RMSE of the estimator is in magnitude. 

Results are presented in Table~\ref{tab:simul1_v1} and~\ref{tab:simul1_v1homo} for different combinations of cross-sectional units and time length, and under different degrees of heteroskedacity. Under heteroskedasticity, the RMSE of PHC0 estimator is much higher than those of the other three estimators for all combinations of $N$ and $T$. The RMSE of the three conservative estimators gradually converges to zero  in large samples, displaying similar values in small samples. Under homoskedasticity, the RMSE of all estimators are always very close to zero for different combinations of panel sample size. The only exception is for $n\le100$ when PHC6 has the smallest RMSE.

%%%%%%%%%%%%%%%%%%%%%%%%%%%%%%%%%%%%%%%%
\subsection{Adjusted Power Test}\label{sec:power}
The power of the test is the average frequency at which the false null hypothesis is rejected in a simulation. In a two-sided single coefficient test, the adjusted power for the false null hypothesis is obtained through the  steps below.
%%%%%%%%%%%%
% 2sided - single test %
%%%%%%%%%%%%
First, for each combination of $\{N,T\}$ and for each simulation run $r = 1,\dots,R$,  compute the test statistics under the false null hypothesis as
\begin{equation*}
T_{N,T}^1(\bfe_{N,T,r})= \frac{\bigl(\bfe_{N,T,r} - \bbeta^1\bigr)}{\sqrt{\AVar}}\distas{a} t_{(df_r, \,\alpha/2)}.
\end{equation*}
%%%
Second, the indicator $\one\{\cdot\}$ turns on every time that the rejection rule holds
\begin{equation*}
J_{N,T,r}^1(\bfe) \equiv \one\bigl\{T_{N,T}^1(\bfe_{N,T,r}) < \textbf{t}_{\alpha/2}^0 \textrm{ or } T_{N,T}^1(\bfe_{N,T,r}) > \textbf{t}_{1-\alpha/2}^0 \bigr\}, 
\end{equation*}  
\noindent where  $\textbf{t}_{\alpha/2}^0$ and $\textbf{t}_{1-\alpha/2}^0$ are values lying respectively at the $(\alpha/2)\th$ and \mbox{$(1-\alpha/2)\th$} percentiles of $T_{N,T}^0(\bfe_{N,T,r})$, and used as critical values\footnote{We cannot use conventional critical values from the t-distribution because size-unadjusted power curves make any comparison between estimators meaningless.}. The empirical critical values differ due to the asymmetric distribution of the test statistics.
Third, count the total number of rejections in the simulation and divide by the number of runs; the adjusted power of a test is 
\begin{equation*}
\bar{J}^1_{N,T,r}(\bfe) \equiv R\inv \sum_{r=1}^R J_{N,T,r}^1(\bfe) = 1-\theta_{test}.
\end{equation*}

%%%%%%%%%%%%
% 2sided - joint test %
%%%%%%%%%%%%
Similarly, for a two-sided test with $q$ linear restrictions the adjusted power of a test is conducted as follows.
First, for each combination of $\{N,T\}$ and for each simulation run $r = 1,\dots,R$,  compute the Wald statistics under the true null hypothesis, $H_0: \mathbf{R}\bm{\beta}-\mathbf{r}^1=\bm{0}$, 
\begin{equation*}
W_{N,T,r}^1(\bfe) = N(\mathbf{R}\bfe_{N,T,r}-\mathbf{r}^1)' \Bigl\{ \mathbf{R}\AVar \mathbf{R}'\Bigr\}\inv ( \mathbf{R}\bfe_{N,T,r}-\mathbf{r}^1) \distas{a}\chi^2(q),
\end{equation*}
\noindent where $\mathbf{r}^1$ is a $q\times1$ vector.
%%%
Second, define the F statistics $F_{N,T,r}^1(\bfe) = W_{N,T,r}^1(\bfe)/q$ under the false null hypothesis for replication run $r$, and sample combination $\{N,T\}$. The rejection rule is defined as
\begin{equation*}
\tilde{J}^1_{N,T,r}(\bfe) \equiv \one\bigl\{F_{N,T,r}^1(\bfe)  > F_{\alpha}^0 \bigr\}, 
\end{equation*}
\noindent where $F^0_{\alpha}$ is the value lying at the $\alpha\th$ quantile of distribution of $F_{N,T,r}^0(\bfe)$  derived under the true null hypothesis, and used as empirical critical in the rejection rule.
%%%
Third, the percentage of rejections that occur in the Monte Carlo exercise is the adjusted power of a test,
\begin{equation*}
\bar{\tilde{J}}^1_{N,T,r}(\bfe) \equiv R\inv \sum_{r=1}^R \tilde{J}^1_{N,T,r}(\bfe) = 1-\theta_{test}.
\end{equation*}

\par  In the simulations, we test $H_0: \beta_j=1$ against $H_1: \beta_j\ne 001$ for $j=\{1,2\}$ for two-sided single coefficient tests, where $\beta^1$ is a value taken from a narrow interval around the true $\beta_j$. For two-sided joint tests we test $H_0: \beta_1=\beta_2=\beta_3=\beta_4=1$  against $H_1: \text{at least one } \beta_j\ne1$,  for $j=1,\dots,4$.

\par
\begin{sidewaysfigure}[p!]
	\caption{Power test for $\beta_1$, heteroskedasticity}\label{fig:power_v1_s1}
    \centering
\begin{minipage}{0.5\textwidth}
	\subfloat{\includegraphics[scale=.85]{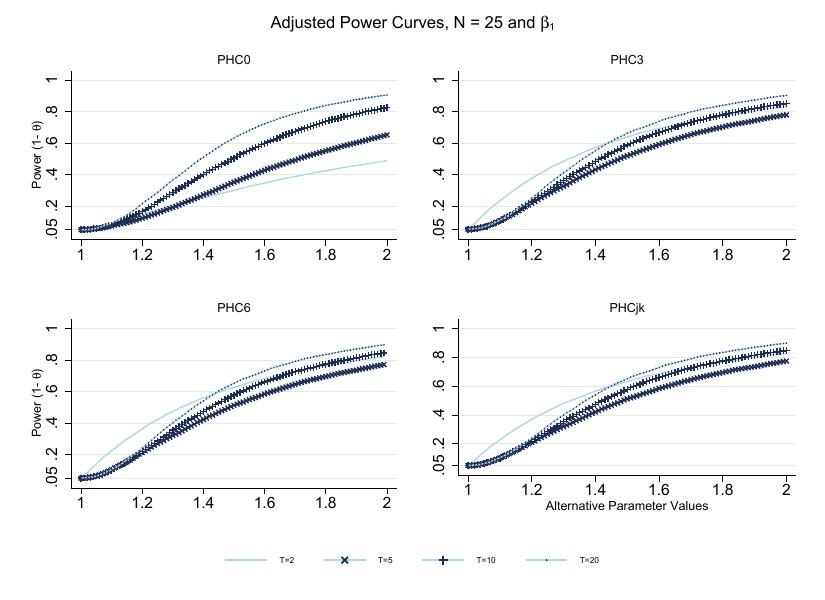}\label{fig:power25_v1_s1}}\\
	\subfloat{\includegraphics[scale=.85]{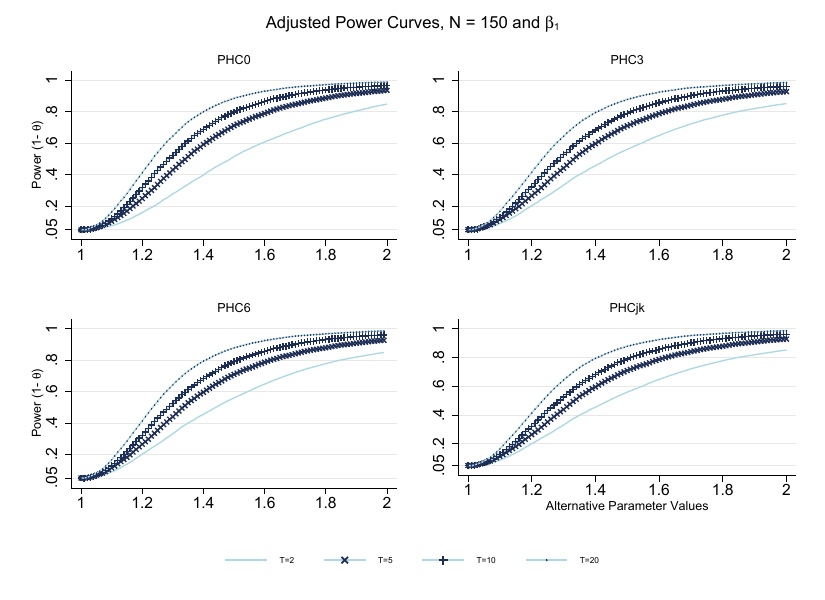}\label{fig:power150_v1_s1}}
\end{minipage}
\begin{minipage}{0.47\textwidth}
	\subfloat{\includegraphics[scale=.85]{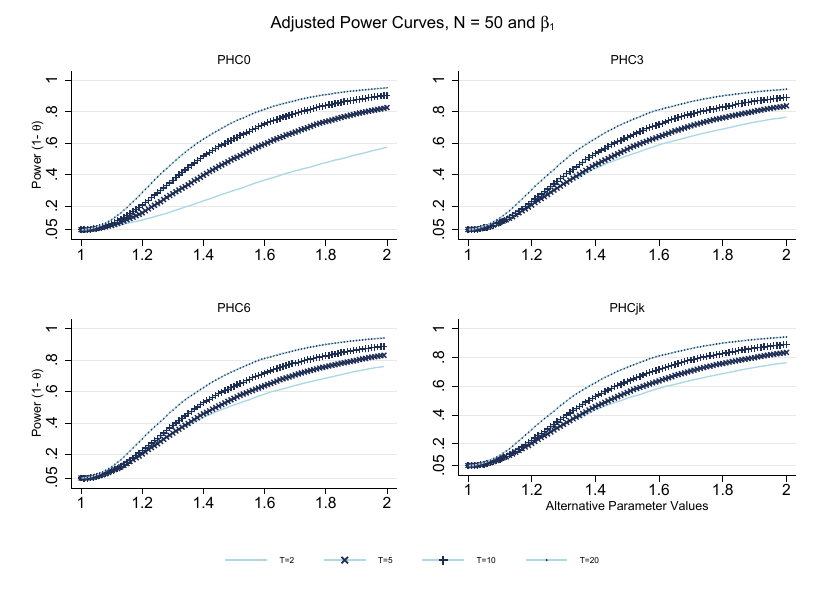}	\label{fig:power50_v1_s1}}\\
	\subfloat{\includegraphics[scale=.85]{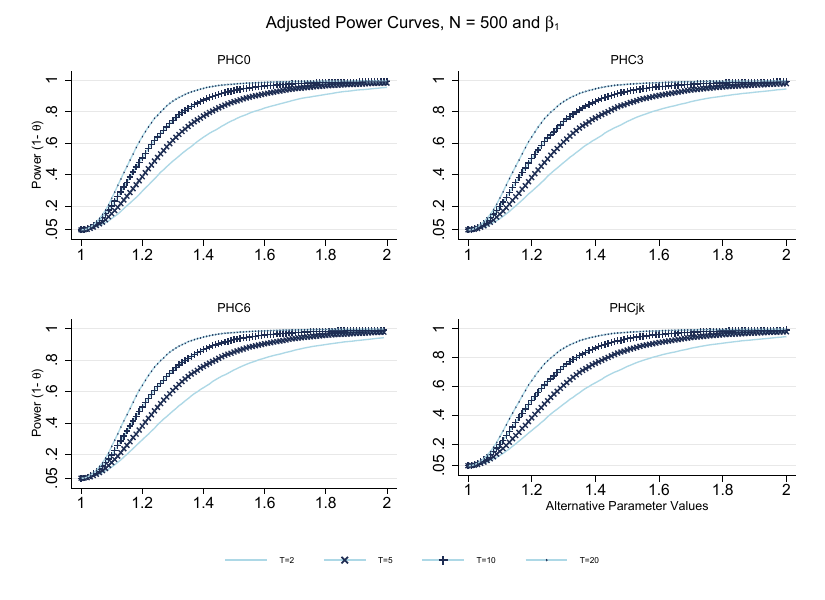}\label{fig:power500_v1_s1}}\\
\end{minipage}
\end{sidewaysfigure}

\begin{sidewaysfigure}[p!]
	\caption{Power test for $\beta_1$, homoskedasticity}\label{fig:power_v1_s1homo}
    \centering
\begin{minipage}{0.5\textwidth}
	\subfloat{\includegraphics[scale=.85]{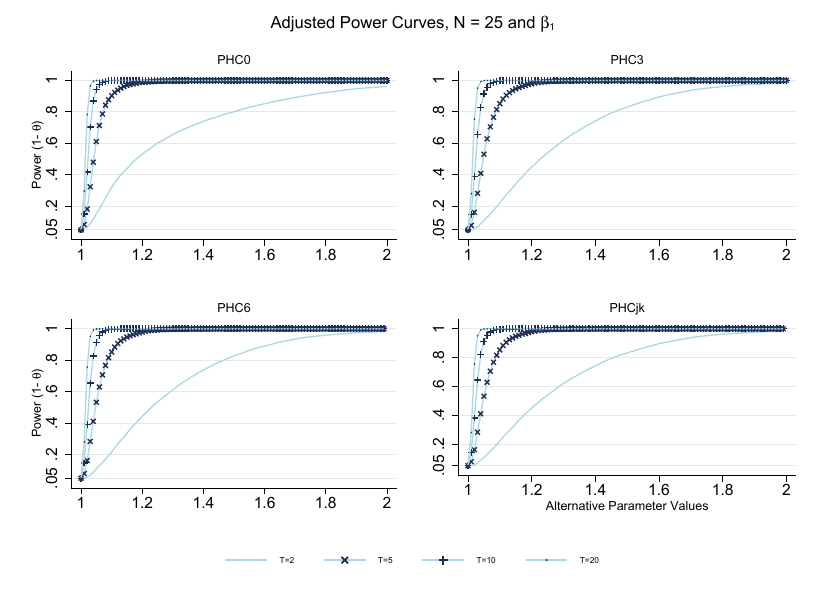}\label{fig:power25_v1_s1homo}}\\
	\subfloat{\includegraphics[scale=.85]{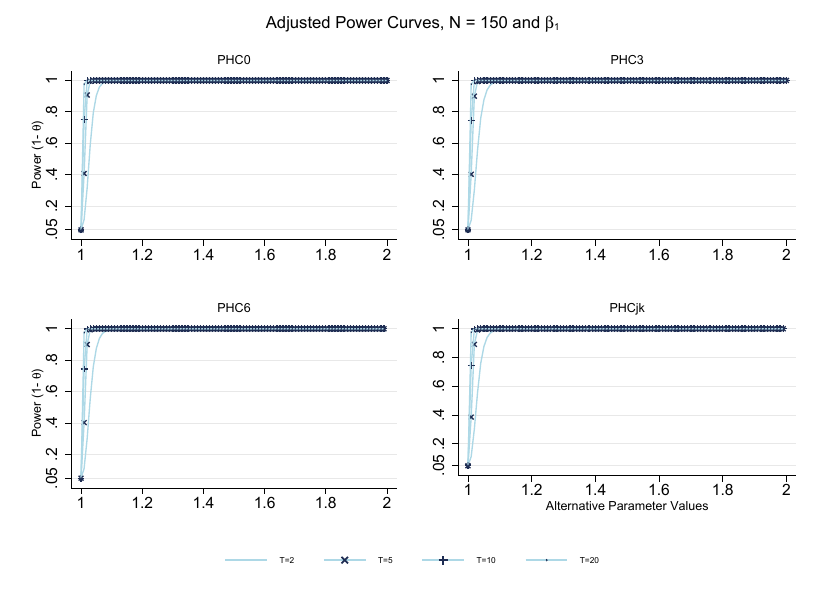}\label{fig:power150_v1_s1homo}}
\end{minipage}
\begin{minipage}{0.47\textwidth}
	\subfloat{\includegraphics[scale=.85]{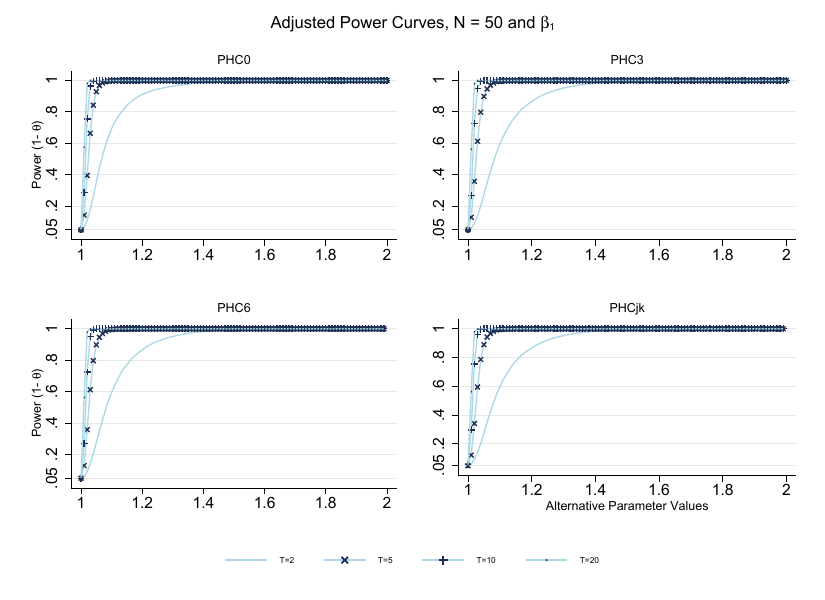}	\label{fig:power50_v1_s1homo}}\\
	\subfloat{\includegraphics[scale=.85]{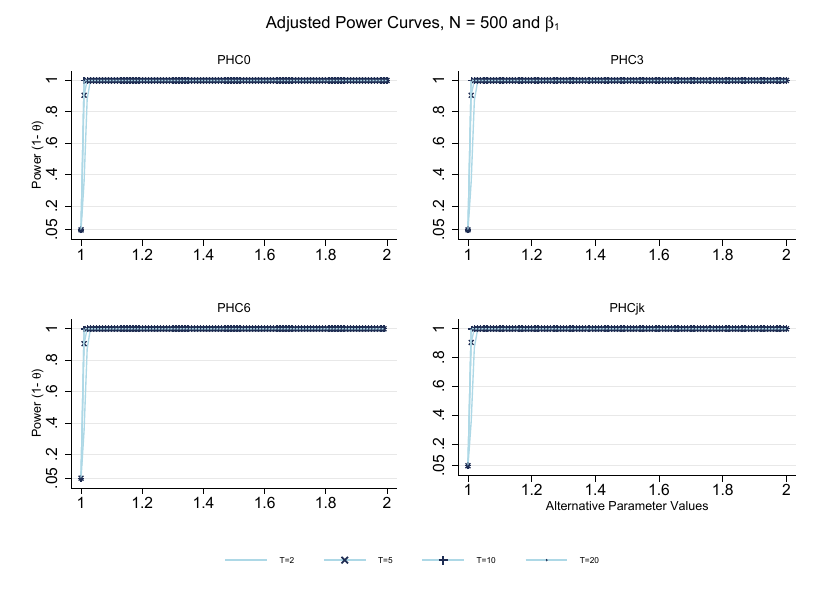}\label{fig:power500_v1_s1homo}}\\
\end{minipage}
\end{sidewaysfigure}

 Figures~\ref{fig:power_v1_s1} and~\ref{fig:power_v1_s1homo} plot size-adjusted power curves of a battery of HC estimators for different panel sample sizes and degree of heteroskedasticity for $\beta_1$. The vertices of all power curves correspond to the nominal size of the test statistics, $\alpha=0.05$. It is common practice to adjust the power for the empirical size because the empirical distributions of test statistics may depend on the nature of the specific regressor and, therefore, any comparison across estimators turns out to be meaningless without size-adjustment. Precisely, in the absence of any size-adjustment the most liberal estimator would tend to have greater power than the most conservative estimator because the former is more likely to over-reject the null hypothesis in favour of the alternative, while the opposite is true for the latter. Unlike the test size, simulation results for the test power do not differ considerably in terms of the overall pattern, but they do in terms of magnitudes. %As we are interested in the global behaviour of the estimators when rejecting a false null hypothesis, we will comment on the results of the two simultion designs together.

Under heteroskedasticity, simulation results show that  PHC0 does not have as good power performance as PHC3, PHC$jk$ and PHC6  in small samples ($N=\{25,50\}$ and especially with $T=2$). In fact, its rejection probabilities at a given parameter value are lower than those of the other three estimators. Fixing T and letting N change, the power performance of PHC0 does not improve. Rejection probabilities remain the lowest and slowly converge to one, even when the distance from the true value of $\beta$ increases. Conversely, we do not observe such a remarkable loss in power when we let $T$ increase and fixing $N$ as the difference with other estimators in the rejection probabilities at a given parameter value becomes negligible or vanishes completely. 

Under the assumption of homoskedasticity, PHC0 has better power than PHC3, PHC$jk$, and PHC6 with $N=25$ for all $T$. This result is in stark contrast with PHC0 poor test size (i.e., RP) described above due to the usual trade-off between type I and type II error. When $T=2$, all power curves show a lower convergence to one.
%In simulation of type 2, power curves of the four estimators computed are well-behaved and approach to the rejection probability of 1 quickly as the distance between the estimated and the tested parameters increases, showing that all estimators have excellent power performances with and without contamination. 
%In general, PHC3, PHC$jk$, and PHC6 are as powerful as PHC0 but they also have better size performances.

\par
\begin{sidewaysfigure}[p!]
	\caption{Power test for joint coefficient test, heteroskedasticity}\label{fig:jpower_s1}
    \centering
\begin{minipage}{0.5\textwidth}
	\subfloat{\includegraphics[scale=.85]{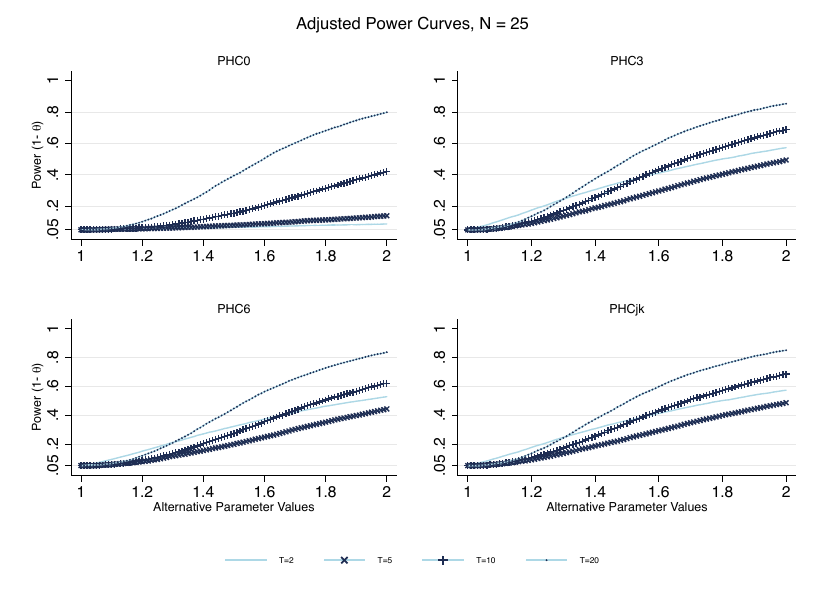}\label{fig:}}\\
	\subfloat{\includegraphics[scale=.85]{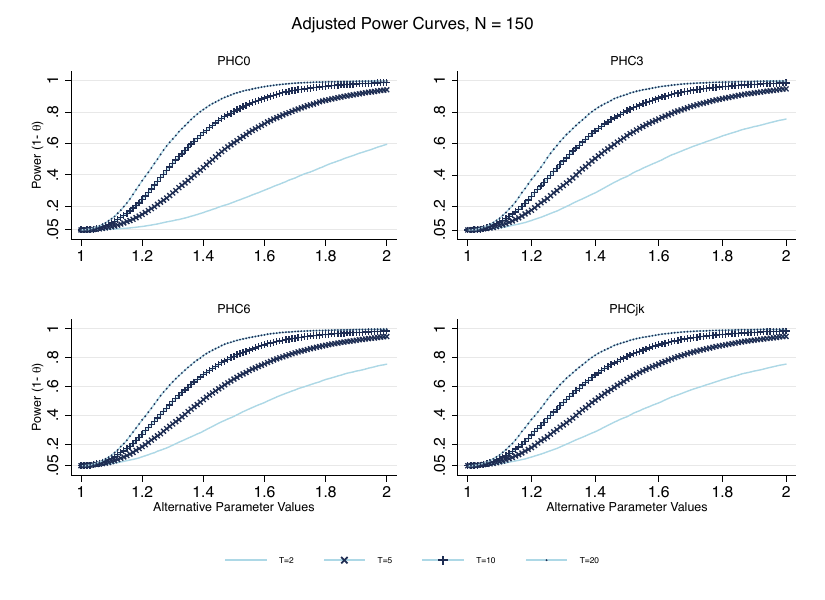}\label{fig:}}
\end{minipage}
\begin{minipage}{0.47\textwidth}
	\subfloat{\includegraphics[scale=.85]{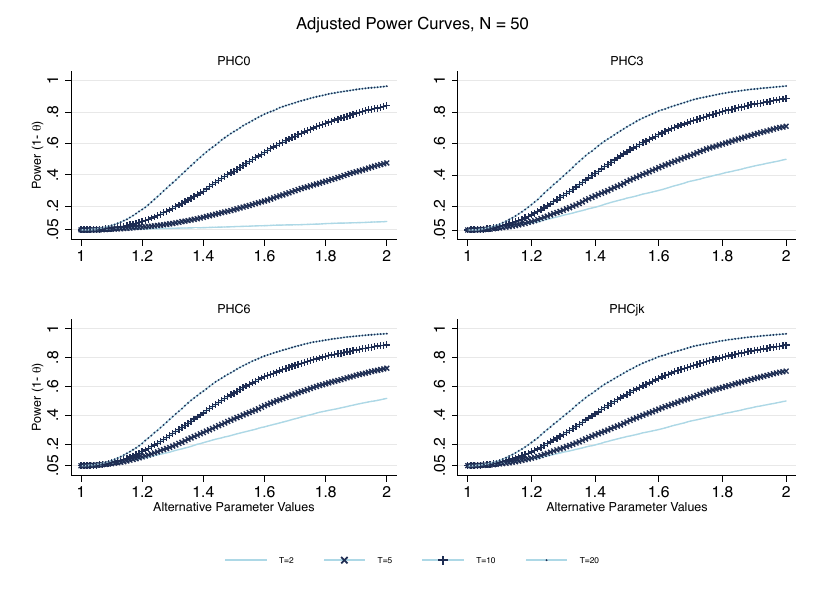}	\label{fig:}}\\
	\subfloat{\includegraphics[scale=.85]{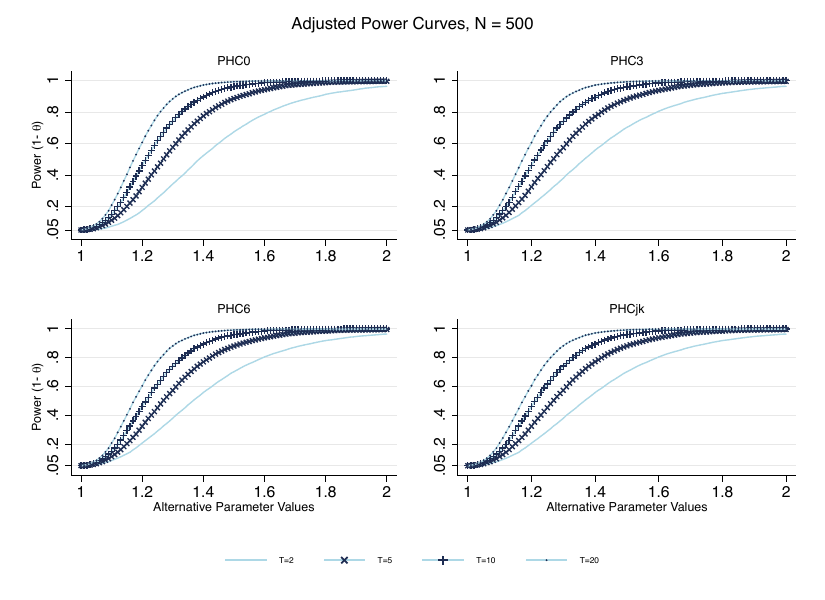}\label{fig:}}\\
\end{minipage}
\end{sidewaysfigure}

\begin{sidewaysfigure}[p!]
	\caption{Power test for joint coefficient test, homoskedasticity}\label{fig:jpower_s1homo}
    \centering
\begin{minipage}{0.5\textwidth}
	\subfloat{\includegraphics[scale=.85]{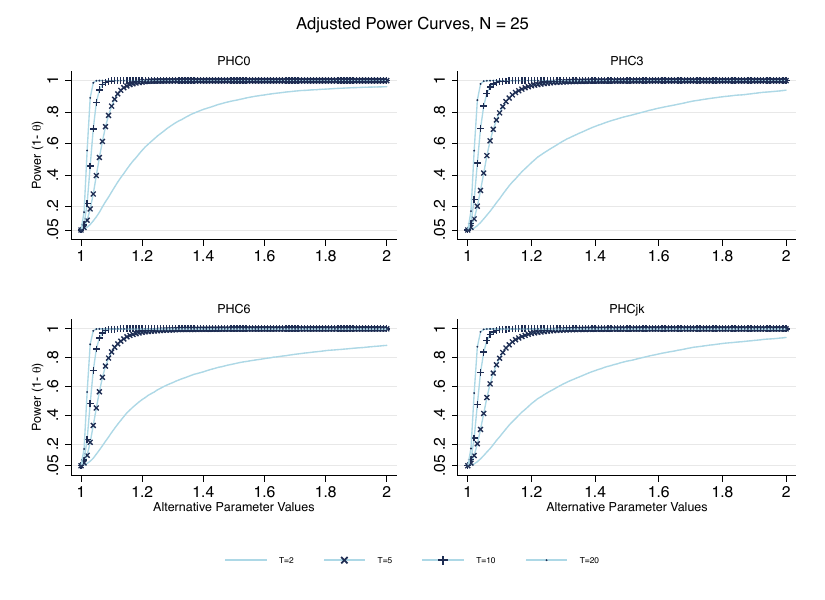}\label{fig:}}\\
	\subfloat{\includegraphics[scale=.85]{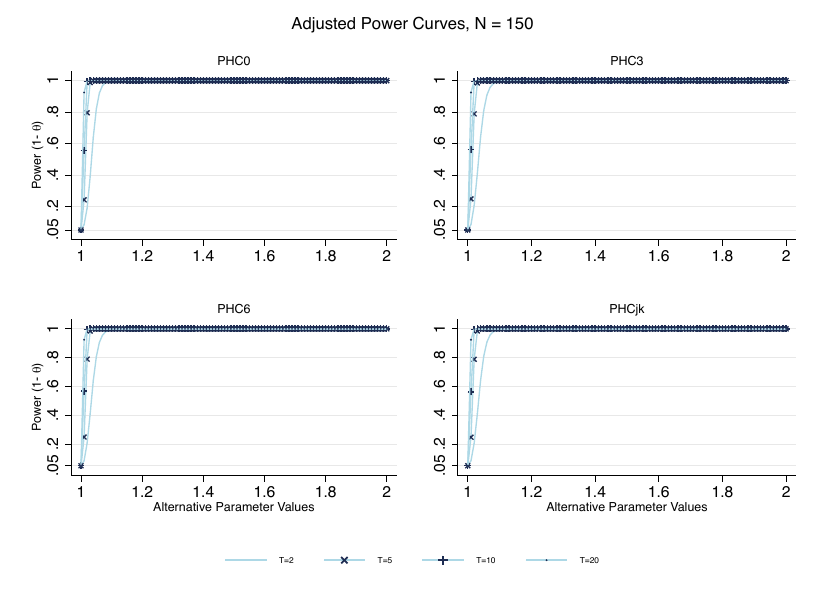}\label{fig:}}
\end{minipage}
\begin{minipage}{0.47\textwidth}
	\subfloat{\includegraphics[scale=.85]{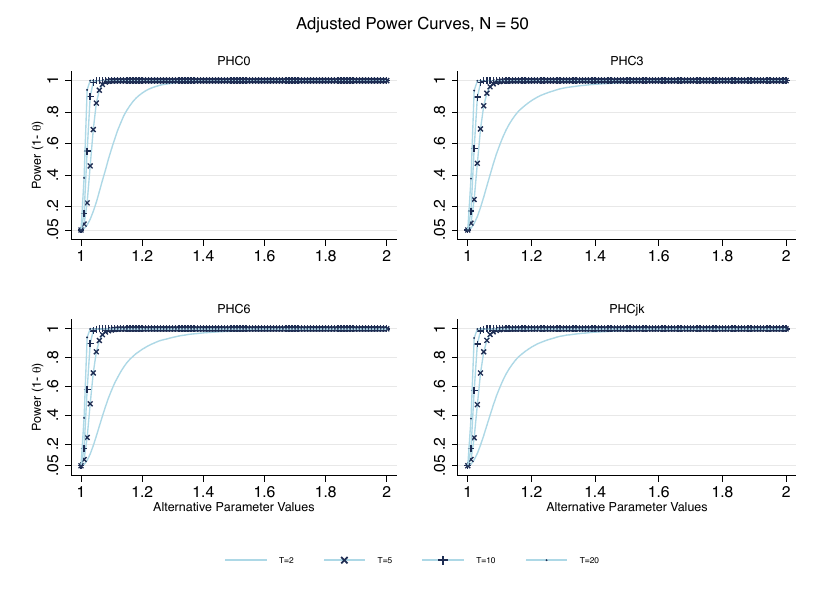}	\label{fig:}}\\
	\subfloat{\includegraphics[scale=.85]{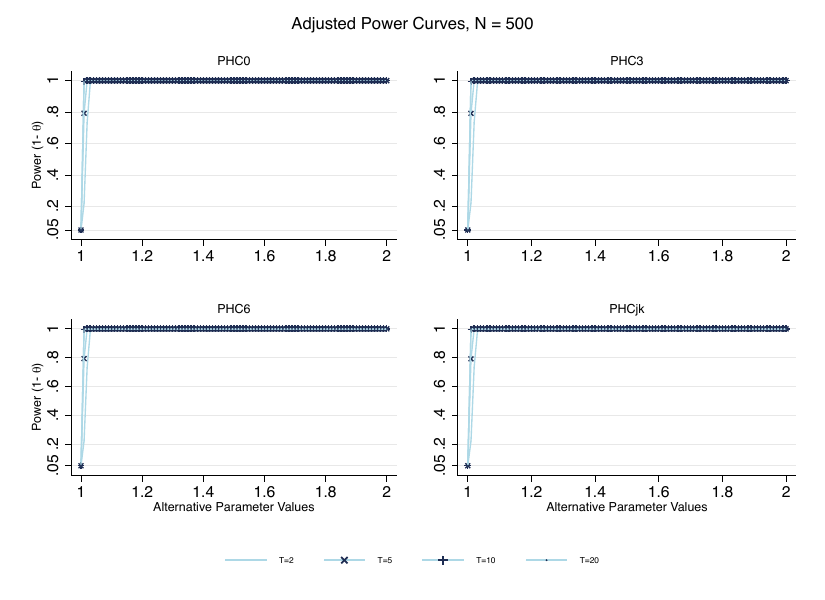}\label{fig:}}\\
\end{minipage}
\end{sidewaysfigure}

 Figures~\ref{fig:jpower_s1} and \ref{fig:jpower_s1homo} show the adjusted power curves  for the joint coefficient test. From the graphs we observe that all power curves are well-behaved under homoskedasticity with rejection rates approaching one quite rapidly as the tested parameter values depart from the true value, and with the increase in the sample size. This cannot be said under heteroskedasticity and, especially, when the panel sample size is small (small $N$ and small $T$) because test statistics of all estimators have low rejection power, especially PHC0 test statistics  when $N=\{25,50\}$ and $T=\{2,5\}$. 

Overall, the four estimators have similar asymptotic behaviour with or without heteroskedasticity. This  can be explained by the sensitivity of the test of hypothesis to sample size. In fact, as the sample size increases the probability of rejecting the false null hypothesis (i.e., the power of the test) increases as well, by construction. The opposite happens to the size of a test instead.

\section{Conclusion}\label{sec:conclusion}
In this chapter, we investigated the effects of the simultaneous presence of a small sample size, heteroskedasticity, and good leveraged data on the validity of conventional statistical inference in linear panel data models with fixed effects. We documented their detrimental effects on the statistical inference calculated with robust standard errors. %that are not powerful enough to reject the false null hypothesis in a narrow interval around the true value.  
More conservative estimators of the sampling variance produce test statistics that have  unbiased empirical sizes and higher power under these circumstance.  %This study provides simulation evidence that these estimators outperform the conventional cluster-robust standard errors under specific circumstances and should be used in linear panel data models as a correction for the statistical inference.  

%%contribution
We formalised a panel version of \citeapos{mackinnon1985} and \citeapos{davidson1993} estimators,  and proposed a new hybrid estimator, $PHC6$. We derived the finite sample properties and the asymptotic distributions of the discussed HC estimators. 
%%what we do
With MC simulations we compared the performances of four types of standard errors, computed with \citeapos{arellano1987} and three types of jackknife-like formulae, in terms of empirical size and power.  We documented the downward bias of conventional robust standard errors under specific circumstances, suggesting alternatives to obtain more reliable statistical inference. 

%%results
The main findings can be summarised as follows. Under heteroskedasticity, more conservative standard errors should be used in the presence of leverage points because their test statistics possess a low proportional bias,  small size distortions, and have higher power. Conversely, conventional  standard errors and the proposed formula, PHC6, should be preferred with homoskedasticity because the other conservative estimators excessively under-reject the true null hypothesis.  Under homoskedasticity cluster-robust formulae should always be used. A similar result was found in \citet{mackinnon1985} and \citet{long2000} for cross-sectional models. 
The cross-sectional dimension matters for the finite sample properties of the estimators but not the size of $N$ relative \mbox{to $T$.} However, conventional cluster-robust standard errors remain upward biased even when their empirical size is correct, and even in larger samples.

%The overall conclusion that can be drawn from the two types of simulation is as follows
%\begin{enumerate}
%\item The cross-sectional dimension -- and not its relative size to the time series dimension -- contributes to the finite-sample behaviour of the estimator;
%\item Without contamination and with heteroskedasticity, PHC3 and PHCjk tend to considerably under-reject the true null hypothesis and are upward biased when the number of cross-sections is small ($N\le50$) but their empirical test size improves as N grows. Conversely, PHC0 and PHC6 are slighlty undersized with few cross-sections independently of the time lenght. However, they tend to be upward biased and, in case of PHC0, its proportional bias is higher in absolute percentage terms than all other estimators.
%\item In case of PHC0, there are cases in which the empirical size is very close to the nominal one, unlike the others, but its proportional bias and RMSE are too large if N is small because the estimated standard errors are extremely liberal.
%\end{enumerate}

%%%%%%%%%%%%%%%%%%%%%%%%%
%%References 
\bibliographystyle{apalike}
\bibliography{biblio.bib}

\begin{thebibliography}{}

\bibitem[Arellano, 1987]{arellano1987}
Arellano, M. (1987).
\newblock Practitioners' corner: Computing robust standard errors for within-groups estimators.
\newblock {\em Oxford bulletin of Economics and Statistics}, 49(4):431--434.

\bibitem[Banerjee and Frees, 1997]{banerjee1997}
Banerjee, M. and Frees, E.~W. (1997).
\newblock Influence diagnostics for linear longitudinal models.
\newblock {\em Journal of the American Statistical Association}, 92(439):999--1005.

\bibitem[Belotti and Peracchi, 2020]{belotti2020}
Belotti, F. and Peracchi, F. (2020).
\newblock Fast leave-one-out methods for inference, model selection, and diagnostic checking.
\newblock {\em The Stata Journal}, 20(4):785--804.

\bibitem[Bertrand et~al., 2004]{bertrand2004}
Bertrand, M., Duflo, E., and Mullainathan, S. (2004).
\newblock How much should we trust differences-in-differences estimates?
\newblock {\em The Quarterly journal of economics}, 119(1):249--275.

\bibitem[Bramati and Croux, 2007]{bramati2007}
Bramati, M.~C. and Croux, C. (2007).
\newblock Robust estimators for the fixed effects panel data model.
\newblock {\em The econometrics journal}, 10(3):521--540.

\bibitem[Cameron et~al., 2011]{cameron2011}
Cameron, A.~C., Gelbach, J.~B., and Miller, D.~L. (2011).
\newblock Robust inference with multiway clustering.
\newblock {\em Journal of Business \& Economic Statistics}, 29(2):238--249.

\bibitem[Cameron and Trivedi, 2005]{cameron2005}
Cameron, A.~C. and Trivedi, P.~K. (2005).
\newblock {\em Microeconometrics: methods and applications}.
\newblock Cambridge university press.

\bibitem[Cattaneo et~al., 2019]{cattaneo2019}
Cattaneo, M.~D., Jansson, M., and Ma, X. (2019).
\newblock Two-step estimation and inference with possibly many included covariates.
\newblock {\em The Review of Economic Studies}, 86(3):1095--1122.

\bibitem[Cattaneo et~al., 2018]{cattaneo2018}
Cattaneo, M.~D., Jansson, M., and Newey, W.~K. (2018).
\newblock Inference in linear regression models with many covariates and heteroscedasticity.
\newblock {\em Journal of the American Statistical Association}, 113(523):1350--1361.

\bibitem[Chesher and Jewitt, 1987]{chesher1987}
Chesher, A. and Jewitt, I. (1987).
\newblock The bias of a heteroskedasticity consistent covariance matrix estimator.
\newblock {\em Econometrica: Journal of the Econometric Society}, pages 1217--1222.

\bibitem[Cribari-Neto, 2004]{cribari2004}
Cribari-Neto, F. (2004).
\newblock Asymptotic inference under heteroskedasticity of unknown form.
\newblock {\em Computational Statistics \& Data Analysis}, 45(2):215--233.

\bibitem[Cribari-Neto and da~Silva, 2011]{cribari2011}
Cribari-Neto, F. and da~Silva, W.~B. (2011).
\newblock A new heteroskedasticity-consistent covariance matrix estimator for the linear regression model.
\newblock {\em AStA Advances in Statistical Analysis}, 95(2):129--146.

\bibitem[Cribari-Neto et~al., 2007]{cribari2007}
Cribari-Neto, F., Souza, T.~C., and Vasconcellos, K.~L. (2007).
\newblock Inference under heteroskedasticity and leveraged data.
\newblock {\em Communications in Statistics - Theory and Methods}, 36(10):1877--1888.

\bibitem[Davidson et~al., 1993]{davidson1993}
Davidson, R., MacKinnon, J.~G., et~al. (1993).
\newblock Estimation and inference in econometrics.
\newblock {\em OUP Catalogue}.

\bibitem[Efron, 1982]{efron1982}
Efron, B. (1982).
\newblock {\em The jackknife, the bootstrap, and other resampling plans}, volume~38.
\newblock Siam.

\bibitem[Eicker, 1967]{eicker1967}
Eicker, F. (1967).
\newblock Limit theorems for regressions with unequal and dependent errors.
\newblock In {\em Proceedings of the fifth Berkeley symposium on mathematical statistics and probability}, volume~1, pages 59--82.

\bibitem[Freedman and Peters, 1984]{freedman1984}
Freedman, D.~A. and Peters, S.~C. (1984).
\newblock Bootstrapping an econometric model: Some empirical results.
\newblock {\em Journal of Business \& Economic Statistics}, 2(2):150--158.

\bibitem[Godfrey, 2006]{godfrey2006}
Godfrey, L. (2006).
\newblock Tests for regression models with heteroskedasticity of unknown form.
\newblock {\em Computational Statistics \& Data Analysis}, 50(10):2715--2733.

\bibitem[Hansen, 2019]{hansen2019}
Hansen, B.~E. (2019).
\newblock Econometrics.
\newblock Unpublished manuscript. Latest version: February 2019.

\bibitem[Hansen, 2007]{hansen2007}
Hansen, C.~B. (2007).
\newblock Asymptotic properties of a robust variance matrix estimator for panel data when t is large.
\newblock {\em Journal of Econometrics}, 141(2):597--620.

\bibitem[Hayes and Cai, 2007]{hayes2007}
Hayes, A.~F. and Cai, L. (2007).
\newblock Using heteroskedasticity-consistent standard error estimators in ols regression: An introduction and software implementation.
\newblock {\em Behavior research methods}, 39(4):709--722.

\bibitem[Hendry, 1984]{hendry1984}
Hendry, D.~F. (1984).
\newblock Monte carlo experimentation in econometrics.
\newblock {\em Handbook of econometrics}, 2:937--976.

\bibitem[Hinkley, 1977]{hinkley1977}
Hinkley, D.~V. (1977).
\newblock Jackknifing in unbalanced situations.
\newblock {\em Technometrics}, 19(3):285--292.

\bibitem[Horn et~al., 1975]{horn1975}
Horn, S.~D., Horn, R.~A., and Duncan, D.~B. (1975).
\newblock Estimating heteroscedastic variances in linear models.
\newblock {\em Journal of the American Statistical Association}, 70(350):380--385.

\bibitem[Huber et~al., 1967]{huber1967}
Huber, P.~J. et~al. (1967).
\newblock The behavior of maximum likelihood estimates under nonstandard conditions.
\newblock In {\em Proceedings of the fifth Berkeley symposium on mathematical statistics and probability}, volume~1, pages 221--233. University of California Press.

\bibitem[Kezdi, 2003]{kezdi2003}
Kezdi, G. (2003).
\newblock Robust standard error estimation in fixed-effects panel models.
\newblock {\em Available at SSRN 596988}.

\bibitem[Kiviet et~al., 2012]{kiviet2012}
Kiviet, J.~F. et~al. (2012).
\newblock {\em Monte Carlo simulation for econometricians}.
\newblock now publishers.

\bibitem[Long and Ervin, 2000]{long2000}
Long, J.~S. and Ervin, L.~H. (2000).
\newblock Using heteroscedasticity consistent standard errors in the linear regression model.
\newblock {\em The American Statistician}, 54(3):217--224.

\bibitem[MacKinnon, 2013]{mackinnon2013}
MacKinnon, J.~G. (2013).
\newblock Thirty years of heteroskedasticity-robust inference.
\newblock In {\em Recent advances and future directions in causality, prediction, and specification analysis}, pages 437--461. Springer.

\bibitem[MacKinnon and White, 1985]{mackinnon1985}
MacKinnon, J.~G. and White, H. (1985).
\newblock Some heteroskedasticity-consistent covariance matrix estimators with improved finite sample properties.
\newblock {\em Journal of econometrics}, 29(3):305--325.

\bibitem[Silva, 2001]{silva2001}
Silva, J.~S. (2001).
\newblock Influence diagnostics and estimation algorithms for powell's scls.
\newblock {\em Journal of Business \& Economic Statistics}, 19(1):55--62.

\bibitem[{\c{S}}im{\c{s}}ek and Orhan, 2016]{csimcsek2016}
{\c{S}}im{\c{s}}ek, E. and Orhan, M. (2016).
\newblock Heteroskedasticity-consistent covariance matrix estimators in small samples with high leverage points.
\newblock {\em Theoretical Economics Letters}, 6(04):658.

\bibitem[Stock and Watson, 2008]{stock2008}
Stock, J.~H. and Watson, M.~W. (2008).
\newblock Heteroskedasticity-robust standard errors for fixed effects panel data regression.
\newblock {\em Econometrica}, 76(1):155--174.

\bibitem[Verardi and Croux, 2009]{verardi2009}
Verardi, V. and Croux, C. (2009).
\newblock Robust regression in stata.
\newblock {\em The Stata Journal}, 9(3):439--453.

\bibitem[Verbeek, 2008]{verbeek2008}
Verbeek, M. (2008).
\newblock {\em A guide to modern econometrics}.
\newblock John Wiley \& Sons.

\bibitem[White, 1980]{white1980}
White, H. (1980).
\newblock A heteroskedasticity-consistent covariance matrix estimator and a direct test for heteroskedasticity.
\newblock {\em Econometrica: Journal of the Econometric Society}, pages 817--838.

\bibitem[White, 1984]{white1984}
White, H. (1984).
\newblock {\em Asymptotic theory for Econometricians}.
\newblock Academic press.

\end{thebibliography}

%%%%%%%%%%%%%%%%%%%%%%%%%%
\appendix
\clearpage
\newpage
\section{Tables and Figures}
%%% T=all, het %%%
%%simul1
\begin{table}[th!]
\centering
%\begin{table}[tp!]
%\centering
\caption{Single hypothesis test, heteroskedasticity}\label{tab:simul1_v1}
\scalebox{.75}{
\begin{threeparttable} 	%use it for notes
   \begin{tabular}{lcccccccccccc}
\vspace{-3mm}\\   
\hline\hline
\vspace{-3mm}\\
&\multicolumn{12}{c}{Heteroskedasticity ($\gamma=2$)}\\
\cmidrule(l{.35cm}r{.25cm}){2-13}\\
\vspace{-8mm}\\
&PB&RP&RMSE&PB&RP&RMSE&PB&RP&RMSE&PB&RP &RMSE \\
$(N,T)$&\multicolumn{3}{c}{(25, 2)}&\multicolumn{3}{c}{(50, 2)}&\multicolumn{3}{c}{(150, 2)}&\multicolumn{3}{c}{(500, 2)}\\
\cmidrule(l{.35cm}r{.25cm}){2-4}\cmidrule(l{.35cm}r{.25cm}){5-7}\cmidrule(l{.35cm}r{.25cm}){8-10}\cmidrule(l{.35cm}r{.25cm}){11-13}\\
\vspace{-8mm}\\		
% \,\, \emph{Estimator}&\multicolumn{12}{c}{}\\		
PHC0&       0.713&0.516&0.305&		  0.625&0.407&0.263&       0.450&0.223&0.147 &       0.316&0.098&0.081\\
PHC3&      -0.083&0.017&0.035&		 -0.138&0.023&0.058&       -0.014&0.029&0.005&       0.084&0.027&0.021\\
PHC6&  -0.042&0.020&0.018&     -0.123&0.025&0.052&       -0.010&0.030&0.003&        0.085&0.027&0.022\\
%PHC6 cv4&  0.285&0.101&0.122&     	-0.084&0.030&0.030&       -0.005&0.031&0.002&        0.085&0.027&0.022\\
PHCjk&      -0.039&0.018&0.017&		 -0.119&0.024&0.050&       -0.010&0.030&0.003 &      0.085&0.027&0.022\\

\vspace{-2mm}\\
&\multicolumn{3}{c}{(25, 5)}&\multicolumn{3}{c}{(50, 5)}&\multicolumn{3}{c}{(150, 5)}&\multicolumn{3}{c}{(500, 5)}\\
\cmidrule(l{.35cm}r{.25cm}){2-4}\cmidrule(l{.35cm}r{.25cm}){5-7}\cmidrule(l{.35cm}r{.25cm}){8-10}\cmidrule(l{.35cm}r{.25cm}){11-13}\\
\vspace{-8mm}\\	
% \,\, \emph{Estimator}&\multicolumn{12}{c}{}\\				
PHC0&0.578&0.337&0.224&		0.473&0.238&0.160&       				0.338&0.112&0.089 &       0.225&0.062&0.042\\
PHC3&-0.121&0.022& 0.047&		-0.024&       0.027&       0.008&       0.066&       0.026&       0.017&       0.099&       0.033&       0.018\\
PHC6&-0.098&0.023& 0.038&			-0.024&0.027&0.008&       0.069&0.026&0.018&       0.100&0.033&0.018\\
%PHC6 cv4 &-0.089&0.025&0.097&			0.034&0.048&0.015&       0.069&0.026&0.018&       0.100&0.033&0.018\\
PHCjk&-0.086 &0.023&0.033&		-0.010&       0.029&       0.003&       0.069&       0.026&       0.018 &       0.100&       0.033&       0.018\\

\vspace{-2mm}\\
&\multicolumn{3}{c}{(25, 10)}&\multicolumn{3}{c}{(50, 10)}&\multicolumn{3}{c}{(150, 10)}&\multicolumn{3}{c}{(500, 10)}\\
\cmidrule(l{.35cm}r{.25cm}){2-4}\cmidrule(l{.35cm}r{.25cm}){5-7}\cmidrule(l{.35cm}r{.25cm}){8-10}\cmidrule(l{.35cm}r{.25cm}){11-13}\\
\vspace{-8mm}\\	
% \,\, \emph{Estimator}&\multicolumn{12}{c}{}\\				
PHC0& 0.472& 0.219 & 0.024&		0.395&0.154&0.117 &       0.277&0.079&0.062&       0.184&0.052&0.027      \\
PHC3& -0.033& 0.022& 0.011&		0.052&0.027&0.015&       	0.098&0.029&0.022&       0.105&0.035&0.015   \\
PHC6&-0.012&0.025& 0.004&		0.061&0.029&0.018&       0.101&0.030&0.023&       0.106&0.035&0.015\\
%PHC6 cv4 &-0.010&0.025&0.003&		0.062&0.029&0.018&       0.101&0.030&0.023&       0.106&0.035&0.015\\
PHCjk& -0.006 & 0.024& 0.002&	0.063&0.029&0.019&       	0.101&0.030&0.023&       0.106&0.035&0.015      \\
\vspace{-2mm}\\

&\multicolumn{3}{c}{(25, 20)}&\multicolumn{3}{c}{(50, 20)}&\multicolumn{3}{c}{(150, 20)}&\multicolumn{3}{c}{(500, 20)}\\
\cmidrule(l{.35cm}r{.25cm}){2-4}\cmidrule(l{.35cm}r{.25cm}){5-7}\cmidrule(l{.35cm}r{.25cm}){8-10}\cmidrule(l{.35cm}r{.25cm}){11-13}\\
\vspace{-8mm}\\	
% \,\, \emph{Estimator}&\multicolumn{12}{c}{}\\				
PHC0&0.385&0.138 &0.112&		0.308 &0.088&0.075&       0.211&0.056&0.037 &      0.131&0.052&0.014     \\
PHC3&0.029&0.021&0.008&		0.076&0.024&0.019&       0.096&0.031 &0.017 &       0.084&0.042 &0.009 \\
PHC6&0.049&0.026& 0.014&			0.085&0.027&0.021&       0.099&0.031&0.017&       0.086&0.042&0.009\\
%PHC6 cv4 &0.049&0.026&0.014&			0.085&0.027&0.021&       0.099&0.031&0.017&       0.085&0.042&0.009\\
PHCjk&0.052 &0.025 &0.015&		0.086 &0.026 &0.021&       0.099&0.031&0.017&       0.085&0.042&0.009\\
\hline
	\end{tabular}	
\begin{tablenotes}[para,flushleft]	
\emph{The number of replications is 10,000. The random variable associated with slope parameter $\beta_1$ is contaminated with leverage points and drives heteroskedasticity. 
PB: Proportional Bias. Positive values indicate by how much the standard error underestimates the ``true'' standard error. 
RP: Rejection Probability of 5\%-level t-test on $\beta_1$ (i.e., size of test).
RMSE: Root Mean Squared Error.
}
  \end{tablenotes}
  \end{threeparttable}
  }
%\end{table}
\end{table}

%%%% T=all, homo %%%
%simul1
\begin{table}[th!]

%\begin{table}[tp!]
\caption{Single hypothesis test, homoskedasticity}\label{tab:simul1_v1homo} 
\centering
\scalebox{.75}{
\begin{threeparttable} 	%use it for notes
   \begin{tabular}{lcccccccccccc}    
\vspace{-3mm}\\   
\hline\hline
\vspace{-3mm}\\
&\multicolumn{12}{c}{Homoskedasticity ($\gamma=0$)}\\
\cmidrule(l{.35cm}r{.25cm}){2-13}\\
\vspace{-8mm}\\
&PB&RP&RMSE&PB&RP&RMSE&PB&RP&RMSE&PB&RP &RMSE \\
$(N,T)$&\multicolumn{3}{c}{(25, 2)}&\multicolumn{3}{c}{(50, 2)}&\multicolumn{3}{c}{(150, 2)}&\multicolumn{3}{c}{(500, 2)}\\
\cmidrule(l{.35cm}r{.25cm}){2-4}\cmidrule(l{.35cm}r{.25cm}){5-7}\cmidrule(l{.35cm}r{.25cm}){8-10}\cmidrule(l{.35cm}r{.25cm}){11-13}\\
\vspace{-8mm}\\	
% \,\, \emph{Estimator}&\multicolumn{12}{c}{}\\		
PHC0&       	   0.369&0.204&0.043&		0.348&0.192&0.014&       0.174&0.116&0.002 &         0.058&0.067&0.000\\
PHC3&     	  -0.411&0.028&0.048&		-0.374&0.040&0.015&       -0.140&0.049&0.002&       -0.049&0.046&0.000\\
PHC6&      -0.328&0.036&0.038&		-0.346&0.044&0.014&       -0.131&0.050&0.002&       -0.046&0.046&0.000\\
%PHC6 cv4&      -0.068&0.097&0.008&		-0.287&0.054&0.012&       -0.114&0.054&0.002&       -0.041&0.047&0.000\\
PHCjk&      	  -0.361&0.030&0.042&		-0.353&0.041&0.014&       -0.135&0.049&0.002 &      -0.048&0.046&0.000\\

\vspace{-2mm}\\
&\multicolumn{3}{c}{(25, 5)}&\multicolumn{3}{c}{(50, 5)}&\multicolumn{3}{c}{(150, 5)}&\multicolumn{3}{c}{(500, 5)}\\
\cmidrule(l{.35cm}r{.25cm}){2-4}\cmidrule(l{.35cm}r{.25cm}){5-7}\cmidrule(l{.35cm}r{.25cm}){8-10}\cmidrule(l{.35cm}r{.25cm}){11-13}\\
\vspace{-8mm}\\	
% \,\, \emph{Estimator}&\multicolumn{9}{c}{}\\						
PHC0&       0.324&0.160&0.008&		0.205&0.118&0.002&       0.074&0.075&0.000&     	  0.017&0.055&0.000\\
PHC3&      -0.310&0.040&0.007& 		-0.160&0.050&0.002&       -0.062&0.047&0.000&       -0.027&0.046&0.000\\
PHC6& -0.284&0.045& 0.007&		-0.148&0.052&0.002&       -0.059&0.048&0.000&      	-0.026&0.047&0.000\\
%PHC6 cv4 & -0.264&0.047&0.006&			-0.140&0.054&0.002&     	-0.056&0.049&0.000&      	-0.025&0.047&0.000\\
PHCjk&      -0.273&0.043&0.006&		-0.146&0.052&0.002&       -0.059&0.048&0.000 &       -0.026&0.047&0.000\\

\vspace{-2mm}\\
&\multicolumn{3}{c}{(25, 10)}&\multicolumn{3}{c}{(50, 10)}&\multicolumn{3}{c}{(150, 10)}&\multicolumn{3}{c}{(500, 10)}\\
\cmidrule(l{.35cm}r{.25cm}){2-4}\cmidrule(l{.35cm}r{.25cm}){5-7}\cmidrule(l{.35cm}r{.25cm}){8-10}\cmidrule(l{.35cm}r{.25cm}){11-13}\\
\vspace{-8mm}\\	
PHC0&       0.197&0.113&0.002&		0.125&0.095&0.001 &       0.041&0.066&0.000&        0.025&0.057&0.000      \\
PHC3&      -0.179&0.043&0.002&		-0.071&0.050&0.000&       -0.032&0.050&0.000&       0.002&0.051&0.000   \\
PHC6& -0.156&0.048& 0.002&		-0.060&0.053&0.000&       -0.028&0.051&0.000&       0.003&0.051&0.000\\
%PHC6 cv4 & -0.151&0.049&0.002&			-0.058&0.053&0.000&       -0.028&0.051&0.000&       0.003&0.051&0.000\\
PHCjk&      -0.151&0.047&0.002&		-0.059&0.052&0.000&       -0.028&0.051&0.000&       0.003&0.051&0.000      \\
\vspace{-2mm}\\

&\multicolumn{3}{c}{(25, 20)}&\multicolumn{3}{c}{(50, 20)}&\multicolumn{3}{c}{(150, 20)}&\multicolumn{3}{c}{(500, 20)}\\
\cmidrule(l{.35cm}r{.25cm}){2-4}\cmidrule(l{.35cm}r{.25cm}){5-7}\cmidrule(l{.35cm}r{.25cm}){8-10}\cmidrule(l{.35cm}r{.25cm}){11-13}\\
\vspace{-8mm}\\	
% \,\, \emph{Estimator}&\multicolumn{12}{c}{}\\					
PHC0&       0.114 &0.084&0.001&		0.065&0.068&0.001&     	  0.019&0.053&0.000 &       0.008&0.052&0.000     \\
PHC3&      -0.097&0.047&0.001&		-0.043&0.048&0.001&      	-0.020&0.045&0.000 &       -0.004&0.049&0.000 \\
PHC6& -0.076&0.053& 0.001&		-0.033&0.050&0.000&         -0.016&0.046&0.000&        -0.003&0.049&0.000\\
%PHC6 cv4 & -0.075&0.053&0.001&			-0.033&0.050&0.000&         -0.016&0.046&0.000&        -0.003&0.049&0.000\\
PHCjk&      -0.074&0.050&0.000&		-0.033 &0.050 &0.000&    	 -0.016&0.046&0.000&       -0.003&0.049&0.000\\
\hline
	\end{tabular} 	
\begin{tablenotes}[para,flushleft]	
\emph{The number of replications is 10,000. The random variable associated with slope parameter $\beta_1$ is contaminated with leverage points and drives heteroskedasticity. 
PB: Proportional Bias. Positive values indicate by how much the standard error underestimates the ``true'' standard error. 
RP: Rejection Probability of 5\%-level t-test on $\beta_1$ (i.e., size of test).
RMSE: Root Mean Squared Error.
}
  \end{tablenotes}
  \end{threeparttable}}
%\end{table}
\end{table}

%%%%%%%%%%%%%%%%%%%%%%%%%%%%%%%%%
\begin{table}[th!]
\centering
%\begin{table}[tp!]
%\centering
\caption{Joint hypothesis test, heteroskedasticity}\label{tab:wsimul_het}
\scalebox{.75}{
\begin{threeparttable} 	%use it for notes
   \begin{tabular}{lcccccccc}
\vspace{-3mm}\\   
\hline\hline
\vspace{-3mm}\\
&\multicolumn{8}{c}{Heteroskedasticity ($\gamma=2$)}\\
\cmidrule(l{.35cm}r{.25cm}){2-9}\\
\vspace{-8mm}\\
&Wald Stats&RP&Wald Stats&RP&Wald Stats&RP&Wald Stats&RP\\

$(N,T)$&\multicolumn{2}{c}{(25,2)}&\multicolumn{2}{c}{(50, 2)}&\multicolumn{2}{c}{(150, 2)}&\multicolumn{2}{c}{(500, 2)}\\
\cmidrule(l{.35cm}r{.25cm}){2-3}\cmidrule(l{.35cm}r{.25cm}){4-5}\cmidrule(l{.35cm}r{.25cm}){6-7}\cmidrule(l{.35cm}r{.25cm}){8-9}\\
\vspace{-8mm}\\	

PHC0&3341260.250&0.923&			29613.070&0.820& 	 6.124&0.510&		1.775&0.210\\
PHC3 &117.439& 0.112& 				 9.203&0.150&			 1.513&0.129&		1.105&0.076\\
PHC6&  66.965&0.264&				 9.947&0.281&     		 2.495&0.223&       1.606&0.189\\
PHCjk &124.333&0.115&			  	 9.505&0.154&			 1.525&0.130&		1.107&0.077\\

\vspace{-2mm}\\
&\multicolumn{2}{c}{(25, 5)}&\multicolumn{2}{c}{(50, 5)}&\multicolumn{2}{c}{(150, 5)}&\multicolumn{2}{c}{(500, 5)}\\
\cmidrule(l{.35cm}r{.25cm}){2-3}\cmidrule(l{.35cm}r{.25cm}){4-5}\cmidrule(l{.35cm}r{.25cm}){6-7}\cmidrule(l{.35cm}r{.25cm}){8-9}\\
\vspace{-8mm}\\			
PHC0&1293.329&0.768&			19.074&0.569& 	2.138&0.263&		1.365&0.115\\
PHC3 & 9.351&0.169& 			 1.964&0.140&		1.175&0.093&		1.081&0.060\\
PHC6& 10.219&0.303&				 2.872&0.248&      1.737&0.197&       1.709&0.217\\
PHCjk &10.377&0.178&			 2.019&0.145&		1.184&0.094&		1.084&0.060\\

\vspace{-2mm}\\
&\multicolumn{2}{c}{(25, 10)}&\multicolumn{2}{c}{(50, 10)}&\multicolumn{2}{c}{(150, 10)}&\multicolumn{2}{c}{(500, 10)}\\
\cmidrule(l{.35cm}r{.25cm}){2-3}\cmidrule(l{.35cm}r{.25cm}){4-5}\cmidrule(l{.35cm}r{.25cm}){6-7}\cmidrule(l{.35cm}r{.25cm}){8-9}\\
\vspace{-8mm}\\			
PHC0&16.202&0.568&		3.348&0.355&      1.540&0.153&       1.210&0.076\\
PHC3&2.157&0.152&		1.353&0.110&      1.086&0.065&       1.053&0.044\\
PHC6&3.565&0.256& 		2.107&0.213&      1.639&0.191&       1.864&0.264\\
PHCjk&2.317&0.163& 		1.389&0.113&      1.094&0.067&       1.055&0.045\\
\vspace{-2mm}\\

&\multicolumn{2}{c}{(25, 20)}&\multicolumn{2}{c}{(50, 20)}&\multicolumn{2}{c}{(150, 20)}&\multicolumn{2}{c}{(500, 20)}\\
\cmidrule(l{.35cm}r{.25cm}){2-3}\cmidrule(l{.35cm}r{.25cm}){4-5}\cmidrule(l{.35cm}r{.25cm}){6-7}\cmidrule(l{.35cm}r{.25cm}){8-9}\\
\vspace{-8mm}\\			
PHC0&4.169&0.375&		1.972&0.213&      1.328&0.098&       1.145&0.064\\
PHC3&1.568&0.119&		1.201&0.082&      1.077&0.052&       1.055&0.046\\
PHC6&2.438&0.234& 		1.820&0.202&      1.773&0.230&       2.311&0.390\\
PHCjk&1.658&0.130&		1.228&0.087&      1.085&0.054&       1.057&0.046\\
\hline
	\end{tabular}	
\begin{tablenotes}[para,flushleft]	
\emph{The number of replications is 10,000. Tested hypothesis $H_0: \beta_1=\beta_2=\beta_3=\beta_4=1$. Random variables associated with slope parameters $\beta_1$ and $\beta_3$ are contaminated with leverage points. All random variables drive heteroskedasticity. RP: Rejection Probability of 5\%-level t-test (i.e., size of test).
}
  \end{tablenotes}
  \end{threeparttable}
  }
%\end{table}
\end{table}

\begin{table}[bh!]
\centering
%\begin{table}[tp!]
%\centering
\caption{Joint hypothesis test, homoskedasticity}\label{tab:wsimul_homo}
\scalebox{.75}{
\begin{threeparttable} 	%use it for notes
   \begin{tabular}{lcccccccc}
\vspace{-3mm}\\   
\hline\hline
\vspace{-3mm}\\
&\multicolumn{8}{c}{Homoskedasticity ($\gamma=0$)}\\
\cmidrule(l{.35cm}r{.25cm}){2-9}\\
\vspace{-8mm}\\
&Wald Stats&RP&Wald Stats&RP&Wald Stats&RP&Wald Stats&RP\\
\vspace{-3mm}\\
$(N,T)$&\multicolumn{2}{c}{(25,2)}&\multicolumn{2}{c}{(50, 2)}&\multicolumn{2}{c}{(150, 2)}&\multicolumn{2}{c}{(500, 2)}\\
\cmidrule(l{.35cm}r{.25cm}){2-3}\cmidrule(l{.35cm}r{.25cm}){4-5}\cmidrule(l{.35cm}r{.25cm}){6-7}\cmidrule(l{.35cm}r{.25cm}){8-9}\\
\vspace{-8mm}\\	
PHC0& 118.278&0.724&			12.974&0.575& 		 2.298&0.298&		1.341&0.136\\
PHC3 &1.444& 0.103& 			 1.478&0.132&			 1.235&0.112&		1.098&0.082\\
PHC6&24.939&0.323&				 10.828&0.427&     	 5.985&0.471&      4.638&0.495\\
PHCjk & 1.521&0.109&			  	 1.514&0.136&			 1.244&0.113&		1.100&0.083\\

\vspace{-2mm}\\
&\multicolumn{2}{c}{(25, 5)}&\multicolumn{2}{c}{(50, 5)}&\multicolumn{2}{c}{(150, 5)}&\multicolumn{2}{c}{(500, 5)}\\
\cmidrule(l{.35cm}r{.25cm}){2-3}\cmidrule(l{.35cm}r{.25cm}){4-5}\cmidrule(l{.35cm}r{.25cm}){6-7}\cmidrule(l{.35cm}r{.25cm}){8-9}\\
\vspace{-8mm}\\	
% \,\, \emph{Estimator}&\multicolumn{12}{c}{}\\				
PHC0&8.066&0.522&			 2.845&0.340& 	1.463&0.162&		1.144&0.087\\
PHC3 &1.855&0.155& 			 1.399&0.132&		1.134&0.086&		1.050&0.066\\
PHC6& 10.577&0.470&			 7.094&0.484&      5.191&0.491&       5.164&0.661\\
PHCjk &1.957&0.166&			 1.431&0.136&		1.142&0.088&		1.053&0.066\\

\vspace{-2mm}\\
&\multicolumn{2}{c}{(25, 10)}&\multicolumn{2}{c}{(50, 10)}&\multicolumn{2}{c}{(150, 10)}&\multicolumn{2}{c}{(500, 10)}\\
\cmidrule(l{.35cm}r{.25cm}){2-3}\cmidrule(l{.35cm}r{.25cm}){4-5}\cmidrule(l{.35cm}r{.25cm}){6-7}\cmidrule(l{.35cm}r{.25cm}){8-9}\\
\vspace{-8mm}\\	
% \,\, \emph{Estimator}&\multicolumn{12}{c}{}\\				
PHC0&3.252&0.343&		1.853&0.216&	 1.265&0.110&       1.075&0.069\\
PHC3&1.588&0.139&		1.286&0.110&      1.102&0.078&       1.026&0.058\\
PHC6&8.413&0.492& 		5.917&0.476&      4.834&0.545&       6.700&0.845\\
PHCjk&1.665&0.147& 		1.313&0.115&      1.109&0.079&       1.028&0.059\\
\vspace{-2mm}\\

&\multicolumn{2}{c}{(25, 20)}&\multicolumn{2}{c}{(50, 20)}&\multicolumn{2}{c}{(150, 20)}&\multicolumn{2}{c}{(500, 20)}\\
\cmidrule(l{.35cm}r{.25cm}){2-3}\cmidrule(l{.35cm}r{.25cm}){4-5}\cmidrule(l{.35cm}r{.25cm}){6-7}\cmidrule(l{.35cm}r{.25cm}){8-9}\\
\vspace{-8mm}\\	
% \,\, \emph{Estimator}&\multicolumn{12}{c}{}\\				
PHC0&2.104&0.227&		1.465&0.143&      1.152&0.079&       1.055&0.066\\
PHC3&1.437&0.121&		1.196&0.088&      1.067&0.060&       1.029&0.059\\
PHC6&6.842&0.234& 		5.376&0.508&      5.584&0.705&       10.257&0.986\\
PHCjk&1.501&0.489&		1.221&0.094&      1.075&0.062&       1.031&0.060\\
\hline
	\end{tabular}	
\begin{tablenotes}[para,flushleft]	
\emph{The number of replications is 10,000. Tested hypothesis $H_0: \beta_1=\beta_2=\beta_3=\beta_4=1$. Random variables associated with slope parameters $\beta_1$ and $\beta_3$ are contaminated with leverage points. All random variables drive heteroskedasticity. RP: Rejection Probability of 5\%-level t-test (i.e., size of test).
}
  \end{tablenotes}
  \end{threeparttable}
  }
%\end{table}
\end{table}
\newpage
%%HET - SIM1

%%HOMO - SIM1

 %%%%%%%%%%%%%%%%%%%%%%%%%%%%%%%%%%%%%
  %%jpower simul1

\clearpage
\newpage
%%%%%%%%%%%%%%%%%%%%%%%%%%%%%%%%%%%%%%%%%
\section{Leave-One-Out (L1O) Estimator}\label{app:l1o}
Following the L1O estimator for RE models in \citet{banerjee1997}, we derive $\bi$ using Woodbury's formula $(\A+\B\D\C)\inv=\A\inv - \A\inv\B \,(\D\inv+\C\A\inv\B)\inv \C\A\inv$, where $\A=\X'\X$, $\B=-\Xii'$, $ \C= \Xii$, and $\D=\I_T$.
\begin{align}
\bi 	& = \biggl(\X'\X-\Xii'\Xii\biggr)\inv \biggl(\X'\Y - \Xii' \yi\biggr)\notag\\
	& =\Biggr(\bigl(\X'\X\bigr)\inv +\bigl(\X'\X\bigr)\inv \Xii' \biggl(\underbrace{\I_T-\Xii \bigl(\X'\X\bigr)\inv \Xii'}_{=\, \I_T- \,\Hi} \biggr)\inv\Xii\bigl(\X'\X\bigr)\inv\Biggr)  \times \biggl(\X'\Y -\Xii' \yi\biggr)\notag\\
	& = \underbrace{\bigl(\X'\X\bigr)\inv\X'\Y}_{=\, \bfe}- \bigl(\X'\X\bigr)\inv\Xii' \yi +\bigl(\X'\X\bigr)\inv\Xii' (\I_T-\Hi)\inv \Xii\underbrace{\bigl(\X'\X\bigr)\inv\X'\Y}_{=\,\bfe}\notag\\
	& - \bigl(\X'\X\bigr)\inv \Xii' (\I_T-\Hi)\inv\underbrace{\Xii \bigl(\X'\X\bigr)\inv\Xii'}_{= \,\Hi} \yi \notag\\
	& = \bfe- \bigl(\X'\X\bigr)\inv \Xii' (\I_T-\Hi)\inv \bigl[(\I_T-\Hi)\yi -\Xii\bfe+\Hi\yi \bigr]\notag\\
	& = \bfe- \bigl(\X'\X\bigr)\inv\Xii' (\I_T-\Hi)\inv (\yi -\Xii\bfe)\notag\\
	& = \bfe- \bigl(\X'\X\bigr)\inv\Xii'(\I_T-\Hi)\inv \resi\notag\\
	& = \bfe- \bigl(\X'\X\bigr)\inv \Xii' \Mi\inv\resi\label{eq:bi},	
\end{align}
%%%
\noindent where $\Mi\inv=(\I_T-\Hi)\inv$. Result~\eqref{eq:bi} is the L1O estimator for FE in \citet{belotti2020}. The sample mean of \eqref{eq:bi} is
\begin{align}
\bmean \equiv&\frac{1}{N} \sum_{i=1}^N\bi = %
	\bfe - \bigl(\X'\X\bigr)\inv  \frac{1}{N} \sum_{i=1}^N\Xii'\Mi\inv\resi  = \bfe - \bigl(\X'\X\bigr)\inv  \mustar \,, \label{eq:bi_mean}
\end{align}
\noindent where $\frac{1}{N} \sum_{i=1}^N\sum_{t=1}^T\,\bfe = N\,\bfe$, $\mustar=\frac{1}{N} \sum_{i=1}^N\Xii'\Mi\inv\resi$ is a $k\times1$ vector.
%%%%%
Therefore, from \eqref{eq:bi} and~\eqref{eq:bi_mean} we get 
%%%%
\begin{align}
 \bi - \bmean & = %
 \bfe- \bigl(\X'\X\bigr)\inv  \Xii'\Mi\inv\resi - \bfe + \bigl(\X'\X\bigr)\inv  \mustar \notag\\
 & = - \bigl(\X'\X\bigr)\inv  \bigl(\Xii'\Mi\inv\resi -  \bigl(\X'\X\bigr)\inv  \mustar\bigr)
\end{align}

%%%%%%%%%%%%%%%%%%%%%%%%%%%%%%%%%%%%%%%%%
%jack
\section{Derivation of the Jackknife Estimator}
Following the procedure in \citet[][pp.~324--326]{hansen2019}, the jackknife estimator of variance can be computed as
%%%%%%%
\begin{align}
\Vjk 	%
& = \Biggl(\frac{N -1}{N}\biggl) \sum_{i=1}^N \Bigl( \bi - \bmean\Bigr) \Bigl(\bi - \bmean\Bigr)' \\
& =  \Biggl(\frac{N -1}{N}\biggl)\bigl(\X'\X\bigr)\inv  \sum_{i=1}^N \, \Biggl\{  \bigl( \Xii'\Mi\inv\resi -    \mustar\bigr) \bigl( \Xii'\Mi\inv\resi -    \mustar\bigr)' \biggl\} \bigl(\X'\X\bigr)\inv \\
& =  \Biggl(\frac{N -1}{N}\biggl)\bigl(\X'\X\bigr)\inv   \Biggl\{ \sum_{i=1}^N \Xii'\Mi\inv\resi \resi'\Mi\inv\Xii  - N \frac{1}{N} \sum_{i=1}^N   \Xii'\Mi\inv\resi\mustar{'} \notag\\
&  -  N \mustar \frac{1}{N}  \sum_{i=1}^N \resi'\Mi\inv\Xii + N\mustar\mustar{'} \Biggl\}\bigl(\X'\X\bigr)\inv \\
& =  \Biggl(\frac{N -1}{N}\biggl)\bigl(\X'\X\bigr)\inv   \Biggl\{ \sum_{i=1}^N \Xii'\Mi\inv\resi \resi'\Mi\inv\Xii  - N\, \mustar\mustar{'} \Biggl\}\bigl(\X'\X\bigr)\inv \\
%& =  \Biggl(\frac{N -1}{N}\biggl)\bigl(\X'\X\bigr)\inv   \Biggl\{\sum_{i=1}^N \Xii'\Mi\inv\resi \resi'\Mi\inv\Xii   - \frac{1}{N} \Biggl(\sum_{i=1}^N \Xii'\Mi\inv\resi \Biggr) \Biggl(\sum_{i=1}^N\resi'\Mi\inv\Xii \Biggr) \Biggl\}\bigl(\X'\X\bigr)\inv \\
& =  \Biggl(\frac{N -1}{N^2}\biggl)\Biggl(\frac{\X'\X}{N}\Biggr)\inv   \Bigl\{\Vh^3 -  \mustar\mustar{'} \Bigl\}\Biggl(\frac{\X'\X}{N}\Biggr)\inv \label{eq:jack1} \,,
\end{align}
\noindent where $\mustar=\frac{1}{N}\sum_{i=1}^N\Xii'\Mi\inv\resi$.

%%%%%%%%%%%%%%%%%%%%%%%%%%%%%
\section{Proof Consistency of Transformed Residuals}\label{app:trans_res}
We show that $\vi = (\I_T - \Hi)\inv\ui \overset{p}{\to}\uii$. We start from 
\begin{align}
\vi - \ui & =  (\I_T - \Hi)\inv\ui - \ui  \notag \\
& = \bigl((\I_T - \Hi)\inv-\I_T\bigr)\ui \notag \\
& = \bigl((\I_T - \Hi)\inv-\I_T\bigr)\bigl(\uii - \Xii(\bfe- \bbeta )\bigr). \label{eqn:tresi1}
\end{align}
%%%%%%%%%%%%%%%
Using Shwarz Inequality and Triangle Inequality for vectors and matrices (\textsc{B.10} and \textsc{B.13}) in \citet[p.795]{hansen2019}, Equation~\eqref{eqn:tresi1} can be rewritten as
\begin{align} 
\|\vi - \ui \| & %%%
= \bb\bigl((\I_T - \Hi)\inv-\I_T\bigr)\bigl(\uii  -\Xii (\bfe- \bbeta)\bigr) \bb \notag \\
& \le \bb(\I_T - \Hi)\inv-\I_T\bb \bb\bigl(\uii- \Xii (\bfe- \bbeta)\bigr) \bb  \label{eqn:tresi2} 
\end{align}
%%%%%%%%%%%%%%%%%%%%%%%%%%%%%%%%%%%%%%%%
\noindent Using Woodbury's formula $(\A+\B\D\C)\inv=\A\inv - \A\inv\B (\D\inv+\C\A\inv\B)\inv \C\A\inv$ with $\A=\I_T$, $\B=\Xii$, $ \C= \Xii'$, and $\D=\X'\X$, $(\I_T - \Hi)\inv$ can be rewritten as follows
\begin{equation} 
(\I_T - \Hi)\inv=\big(\I_T - \Xii(\X'\X)\inv\Xii'\big)\inv=\I_T + \Xii(\X'\X-\Xii'\Xii)\inv\Xii'
\end{equation}
\noindent and, hence, the first component in \eqref{eqn:tresi2}  
\begin{equation} \label{eqn:Mi} 
\begin{split}
\bb(\I_T - \Hi)\inv-\I_T\bb %%%
& = \bb\Xii(\X'\X-\Xii'\Xii)\inv\Xii'\bb  \\
& \le \bb\Xii\bb^2 \bb\bigl(\X'\X-\Xii'\Xii\bigr)\inv  \bb\\
&= \frac{1}{N} \bb\Xii\bb^2 \,\Bigg\|\biggl( \frac{1}{N}\X'\X- \frac{1}{N}\Xii'\Xii\biggr)\inv  \Bigg\|.
\end{split}
\end{equation}
%%%%%%%%%%%%%%%%%%%%%%%%%%%%%%%%%%%%%%%%
 Then, using expression~\eqref{eqn:Mi} in inequality~\eqref{eqn:tresi2} 
\begin{align}
\|\vi - \ui \| %%%
& \le \frac{1}{N} \bb\Xii\bb^2 \,\Bigg\|\biggl( \frac{1}{N}\X'\X- \frac{1}{N}\Xii'\Xii\biggr)\inv  \Bigg\|\bigl(\|\uii \| + \bb\Xii\bb \bb\bfe- \bbeta\bb\bigr) \label{eqn:tresi3}  \\
&= o_p(1)\notag
\end{align}
%%%%%
\noindent where the first component of~\eqref{eqn:tresi3} is $O_p(N^{-1})$  under \ref{item:mom}.\ref{item:mom_x2} for $r\ge2$; the  second component involves $\SSXX+o_p(1)$ as $N\inv \X'\X\overset{p}{\to}\E(\X'\X)= \SSXX>0$ by the Central Limit Theorem; the last component is $O_p(1)$ because the random variables in parenthesis are $O_p(1)$ under \ref{item:mom}.\ref{item:mom_x2}-\ref{item:mom_Adistr}, and $\bb\bfe- \bbeta \bb\overset{p}{\to}0$ is $o_p(1)$. Therefore, the overall expression is bounded above by an $o_p(1)$ random variable.
%%%%%%%%%%%%%%%%%%%%%%%%%%%%%%%%%
%%%%
 Note that
\begin{equation} \label{eqn:errore} 
\|\ui - \uii \|  %%%
%&\le \max_{1\le i\le N}\bb\Xii (\bfe- \bbeta)\bb  \\
 \le \bb\Xii\bb \bb\bfe- \bbeta\bb
\end{equation}
\noindent because $\Xii$ is $O_p(1)$ by \ref{item:mom}.\ref{item:mom_x2} and $\bb\bfe- \bbeta \bb\overset{p}{\to}0$, $\|\ui - \uii \|$ is $o_p(1)$. Therefore,  $\ui \overset{p}{\to}\uii$ as $N\to\infty$  and $T$  fixed.
%\par Using \textsc{thm}~7.17 in \citet[p.249]{hansen2019} in conjuction with the aforementioned assumption, we get the following result
Using result in~\eqref{eqn:tresi3} and \eqref{eqn:errore}, we obtain the desired result
\begin{equation}\label{eqn:tresi4}
\vi = \ui +o_p(1)\overset{p}{\to}\uii\,\, \text{ as } N\to\infty \text{ and } T \text{ fixed}.
\end{equation}
\noindent Result ~\eqref{eqn:tresi4} shows that the transformed standard errors $\vi$ are a uniformly consistent estimator for the error term $ \uii $. This result guarantees the consistency of any other formula of alternative HC estimators, such as $PHC3$ and $PHC6$. In fact, using Equation~\eqref{eqn:tresi4}  we can show that
%%%%
\begin{align}
\Vhh-\Vh& = \frac{1}{N} \sum_{i=1}^N\Xii'\vi\vi'\Xii - \frac{1}{N} \sum_{i=1}^N\Xii'\ui\ui'\Xii\notag \\
& = \frac{1}{N} \sum_{i=1}^N\Xii'\Big(\vi(\vi'-\ui')(\vi+\ui)\ui'\Big)\Xii
\end{align}
%%%%
\noindent Then,
%%%%
\begin{align}\label{eq:VV}
\Bb\Vhh-\Vh\Bb& \le  \frac{1}{N} \sum_{i=1}^N\bb\Xii'\vi(\vi'-\ui')\Xii\bb +  \frac{1}{N}\sum_{i=1}^N\bb\Xii'(\vi+\ui)\ui'\Xii\bb\notag \\
& \le \max_{1\le i\le N}\|\vi - \ui \| \Biggl( \frac{1}{N}\sum_{i=1}^N\bb\Xii\Xii'\vi\bb + \frac{1}{N} \sum_{i=1}^N\bb\Xii\Xii'\ui'\bb \Biggr)\notag \\
& \le \max_{1\le i\le N}\|\vi - \ui \| \Biggl( \frac{1}{N}\sum_{i=1}^N\bb\Xii\bb^2\bb\vi\bb + \frac{1}{N} \sum_{i=1}^N\bb\Xii\bb^2\bb\ui\bb \Biggr)\\
&  = o_p(1)\notag
\end{align}
%%%%
\noindent where the first term of \eqref{eq:VV} is $o_p(1)$ from result~\eqref{eqn:tresi3}; the two components in parenthesis are the sums of random variables with finite means both converging in probability to $\E\Bigl(\bb\Xii\bb^2\bb\uii\bb \Bigr)$ by  \ref{item:mom} and results~\eqref{eqn:errore}-\eqref{eqn:tresi4} and, therefore, $O_p(1)$. Their product is $o_p(1)$.

%\noindent By \textsc{thm}s~7.6-7.7 in \mbox{\citet[pp.230--232]{hansen2019}} and \ref{item:V}, the desired result follows 

\par The last step left to show is  $\Vh\overset{p}{\to}\Vv$ such that $\Vhh=\Vh+o_p(1)\overset{p}{\to}\Vv$  as  $N\to\infty$  and $ T $ fixed. Following \citet[pp.230--232]{hansen2019}, we start from the definition of conventional robust variance-covariance matrix 
%%%%%%%%%%%%%%%%%%%%%%%%%%%%
\begin{align}\label{eq:Vhat}
\Vh& =  \frac{1}{N} \sum_{i=1}^N \Xii'\ui\ui'\Xii \notag\\
& \frac{1}{N} \sum_{i=1}^N \Xii'\uii\uii'\Xii + \frac{1}{N} \sum_{i=1}^N \Xii'(\ui\ui'-\uii\uii')\Xii
\end{align}
\noindent where the first component of Equation~\eqref{eq:Vhat}  converges in probability to $\E(\Xii'\Sgma\Xii)=\Vi$ by \ref{item:xu}.\ref{item:het}  by \textsc{lie} and \ref{item:dgp} with finite limit $\Vv$ under \textsc{thm~6.16} in \citet[p.189]{hansen2019} for sequences of \emph{inid} random variables, provided that \ref{item:V} and \ref{item:mom}.\ref{item:mom_x2} hold. The second component needs to converge in probability to zero to claim consistency of $\Vh$.
%%%%%%%%%%%%%%%%%%%%%%%%%%%%
%the Triangle Inequality (\textsc{B.14}) in \citet{hansen2019}
Applying  matrix norm to~\eqref{eq:Vhat}, we get 
\begin{align}\label{eq:uu}
\Bigg\|\frac{1}{N} \sum_{i=1}^N \Xii'(\ui\ui'-\uii\uii')\Xii\Bigg\| %%%
& \le \frac{1}{N} \sum_{i=1}^N  \Bb  \Xii'(\ui\ui'-\uii\uii')\Xii\Bb\notag \\
& \le \frac{1}{N} \sum_{i=1}^N  \bb\Xii\bb^2 \bb\ui\ui'-\uii\uii'\bb
\end{align}
%%%%%%%%%%%%%%%%%%%%%%%%%%%%
\noindent Note that
%%%%%
\begin{align}\label{eq:uu2}
\ui\ui' & = \Bigl(\uii - \Xii(\bfe- \bbeta) \Bigr)\Bigl(\uii - \Xii(\bfe- \bbeta) \Bigr)'\notag\\
& = \uii\uii' - \uii(\bfe- \bbeta)' \Xii'- \Xii(\bfe- \bbeta) \uii' +\Xii(\bfe- \bbeta)(\bfe- \bbeta)'\Xii'
\end{align}
%%%%%%%%%%%%%%%%%%%%%%%%%%%%
Rearranging last line of Equation~\eqref{eq:uu2} and using the Triangle Inequality (\textsc{B.14}) and Schwarz Inequality  (\textsc{B.13}), we obtain
%%%%%%%%%%%
\begin{align}\label{eq:uu3}
\bb\ui\ui' - \uii\uii'\bb %%%
%& \le \Bb  \uii(\bfe- \bbeta)' \Xii' +\Xii(\bfe- \bbeta) \uii' +\Xii(\bfe- \bbeta)(\bfe- \bbeta)'\Xii'\Bb\notag \\
& \le 2  \Bb \Xii(\bfe- \bbeta) \uii'\Bb + \bb\Xii\bb^2 \Bb\bfe- \bbeta\Bb^2\notag \\
& \le 2  \bb \Xii\bb \bb\uii\bb\Bb\bfe- \bbeta\Bb + \bb\Xii\bb^2 \Bb\bfe- \bbeta\Bb^2
\end{align}
%%%%%%%%%%%%%%%%%%%%%%%%%%%% 
Plugging \eqref{eq:uu3} in  \eqref{eq:uu}
\begin{align}\label{eq:uu4}
\Bigg\|\frac{1}{N} \sum_{i=1}^N \Xii'(\ui\ui'-\uii\uii')\Xii\Bigg\| %%%
& \le \frac{1}{N} \sum_{i=1}^N  \bb\Xii\bb^2  \bb \ui\ui'-\uii\uii'\bb \notag \\
& \le \frac{1}{N} \sum_{i=1}^N  \bb\Xii\bb^2 \Bigg\{2  \bb \Xii\bb \bb\uii\bb \Bb\bfe- \bbeta\Bb+ \bb\Xii\bb^2 \Bb\bfe- \bbeta\Bb^2\Bigg\}\notag \\
& \le 2 \Bigg(\frac{1}{N} \sum_{i=1}^N  \bb\Xii\bb^3 \bb\uii\bb \Bigg)\Bb\bfe- \bbeta\Bb + \Bigg(\frac{1}{N} \sum_{i=1}^N \bb\Xii\bb^4 \Bigg)\Bb\bfe- \bbeta\Bb^2\notag \\
&=o_p(1)
\end{align}
where the average in the first parenthesis is $O_p(1)$ because it is the mean of random variables bounded above by finite quantities, that is,  $\E\big(\bb\Xii\bb^3 \bb\uii\bb\big)\le \big(\E\bb\Xii\bb^3\big)^{\frac{3}{4}} \big(\E\bb\uii\bb^4\big)^{\frac{1}{4}} $ by evoking H\"{o}lder's Inequality (\textsc{B.28}) in \citet[p.~796]{hansen2019}  and under \ref{item:dgp} and \ref{item:mom}.\ref{item:mom_x2}-\ref{item:mom_Adistr}; the average in the second parenthesis is $O_p(1)$ as the average of a random variable with finite mean by \ref{item:mom}.\ref{item:mom_x2}; $\Bb\bfe- \bbeta\Bb\overset{p}{\to}0$ and, thus, \mbox{is $o_p(1)$.}
%%%%%%%%%%%%%%%%%%%%%%%%%%%% 
It follows that $\Vh\overset{p}{\to}\Vv$ and, therefore, the desired result
\begin{equation}
\Vhh=\Vh+o_p(1)\to\Vv.
\end{equation}

%% Theorem 7.3 in \citet[p.222]{hansen2019} (pdf p.237)
%% Theorem 7.6 in \citet[p.230]{hansen2019} (pdf p.245)
%% Theorem 7.7 in \citet[p.232]{hansen2019} (pdf p.247)
%% Theorem 7.18 in \citet[p.250]{hansen2019} (pdf p.265)
%% Theorem 7.17 in \citet[p.249]{hansen2019} (pdf p.264)

%%%%%%%%%%%%%%%
\section{Consistency of PHC6}

The proposed estimator, PHC6, of the asymptotic variance-covariance matrix is  
\begin{equation}\label{eq:PHC6}
\avar_6=c_6 \, \Sxx\inv\Vh^6\Sxx\inv\,,
\end{equation}
\noindent where the variance-covariance matrix is $\Vh^6 = \frac{1}{N}\sum_{i=1}^N \Xii' \vi\vi'\Xii,$ and the matrix $\Mi$ has functional form 
\begin{equation}\label{eq:Mi}
\Mi =%
\begin{cases}
\I_T  & \text{if }\histar<2\\
\I_T-\Hi & \text{otherwise} 
\end{cases}
\end{equation}
\noindent where $\histar = max\big\{h_{i11}/\overline{h}_{11},\dots,h_{iTT}/\overline{h}_{TT}\big\}$ is the maximal individual leverage of unit $i$; and $\bhtt = N\inv\sum_{i=i}^N\hitt$ is the average leverage at time $t$, with $\hitt$ being the individual leverage of unit $i$ at time $t$. 
 The finite sample correction of PHC6 is
	\begin{equation*}
		 c_6=
		\begin{cases}
		\frac{(NT-1)N}{(NT-k)(N-1)}& \text{if } \histar<2\\
		\frac{N-1}{N}&  \text{otherwise} 
		\end{cases}
\\
	\end{equation*}
\noindent As $N\to\infty$ and $T$ is fixed, $\M = (\I_T - \Hi)$	because the selection criterion is $2<\infty$. As argued above, $\Hi \overset{p}{\to}\zero$ as $N\to\infty$ and $T$ fixed because leverage measures are asymptotically negligible \citep[p.249]{hansen2019}. Therefore, PHC6 collapses to PHC3 that converges to PHC0 which is a consistent estimator of the asymptotic variance \citep[Theorem 7.7, p.232]{hansen2019}. From \citeapos{white1980} general result and under  the above model assumptions and \textsc{thm}~7.7 in \citet[p.232]{hansen2019}, it follows $\widetilde{\bm{\Sigma}} = \widehat{\bm{\Sigma}} +o_p(1)\to \Ssgma$ as $N\to \infty$ and $T$ fixed such that $\avar_6\overset{p}{\to}\mathrm{AVar}(\bfe)$, making PHC6 consistent estimator of the sampling variance.
%\citep[Theorem 7.7, p.232]{hansen2019} (pdf p.247)
%thm 7.6 p.230 (pdf p.245) for consistency HC0
%%%%%%%%%%%%%%%%%%%%%%%%%%%%%%%%%%%%%%%%%
% Derivation distribution of distribution of W
%By the assumptions of \emph{iid} and normality  of $\x$, the random variable $\W$ is normally distributed with  mean $\mu_w = \beta_0+ \sum_{j=1}^J\beta_j \mu_{x_j}$ and variance $\sigma^2_w = \sum_{j=1}^J\beta^2_j \sigma^2_{x_j}$, where $\mu_{x_j} = \E\bigr(x_{it,j}\bigr)$ and \mbox{$\sigma^2_{x_j} = \mathrm{Var}\bigr(x_{it,j}\bigr)$}  when $\gamma=1$. If $\gamma=2$, the random variable $\W^2$ has mean  and variance equal to respectively
%%%%
%\begin{align}
%\mu_{w^2} &  = \beta_0^2+ \sum_{j=1}^J\beta_j^2 \bigl(\sigma^2_{x_j}+\mu^2_{x_j}\bigr)+2\beta_0 \sum_{j=1}^J\beta_j \mu_{x_j}+2\sum_{\substack{k,j=1,\\ k \ne j}}^J\beta_j\beta_j \mu_{x_j}\mu_{x_j} \\
%\sigma_{w^2} & = \sum_{j=1}^J\beta_j^4\sigma^2_{x_j} + 4 \beta_j^2 \sum_{j=1}^J\beta_j^2\sigma^2_{x_j}+ 4 \sum_{\substack{j,k=1, \\j \ne k}}^J\beta_j^2 \beta_k^2 \bigl(\sigma^2_{x_j}\sigma^2_{x_k} + \sigma^2_{x_j}\mu^2_{x_k}+ \sigma^2_{x_j}\mu^2_{x_k}),
%\end{align}
%\noindent where $\mu^2_{x_j} = \big[\E\big(x_{it,k}\big)\big]^2$, $\mu_{x_j} = \E\bigr(x_{it,j}\bigr)$, and  $\sigma^2_{x_j} = \mathrm{Var}\bigr(x_{it,j}\bigr)$. Standardising $\W$, we get $\frac{\W-\mu_w}{\sigma^2_w} \distas{} \N \bigl(0,1\bigr)$ if $\gamma=1$, and $\Bigr(\frac{\W-\mu_w}{\sigma^2_w}\Bigr)^2 \distas{}  \chi^2_1$ if $\gamma=2$. \\

\section{Derivation of the Distribution of $\W$}\label{sec:w}
The error term $\error$ is intrinsically heteroskedastic but not on average due to the presence of the scaling factor $z(\gamma)$. Let $\W=\beta_0+  \sum_{j=1}^J \beta_j x_{it,j}$ with $\{x_{it,j}\}_{j=1}^J$.  When  $\gamma=1$, the mean and variance of a random variable W with an unknown distribution are as follows
\begin{align} \label{eq:mean_w_iid}
\E(\W)%
& = \E\Biggl[\beta_0+  \sum_{j=1}^J \beta_j x_{it,j}\Biggr] %
 = \beta_0+  \sum_{j=1}^J \beta_j \, \E\bigl( x_{it,j}\bigr) %
 = \beta_0+  \sum_{j=1}^J \beta_j \mu_{x_j}\\
 %%%%%%%%%%%%%%%
\mathrm{Var}\bigl(\W\bigl) %
 &= \mathrm{Var}\Biggl[\beta_0+  \sum_{j=1}^J \beta_j x_{it,j}\Biggr]%
 =   \sum_{j=1}^J \beta^2_j \, \mathrm{Var}\bigl( x_{it,j}\bigr) + 2\sum_{\substack{j,k=1\\j \ne k}}^J\beta_j\beta_k\mathrm{Cov}( x_{it,j}, x_{it,k})  \notag\\
 &=   \sum_{j=1}^J \beta^2_j \sigma^2_{x_j} ,\label{eq:var_w_iid}
\end{align}
\noindent where  $\E\bigl( x_{it,j}\bigr)=\mu_{x_j}$,  $\mathrm{Var}\bigl( x_{it,j}\bigr)=\sigma^2_{x_j}$, and $\mathrm{Cov}( x_{it,j}, x_{it,k})=0$ because the independence assumption guarantees that $\E\bigl( x_{it,j}, x_{it,k}\bigr) = \E\bigl( x_{it,j}\bigr) \E\bigl( x_{it,k}\bigr)$. The results are valid under  independent and identically distributed (\emph{iid}) random variables. 
By the assumptions of \emph{iid} and normality  of $\x$, the random variable $\W$ is normally distributed with mean~\eqref{eq:mean_w_iid} and variance~\eqref{eq:var_w_iid}. When the regressors are drawn from a standard normal distribution, \eqref{eq:mean_w_iid} and \eqref{eq:var_w_iid} reduce to $\beta_0$ and $\sum_{j=1}^J \beta^2_j$, respectively. Thus, $\W \distas{} \N \Bigl(\beta_0, \sum_{j=1}^J \beta^2_j \Bigr)$. Standardising $\W$, we get $\frac{\W-\mu_w}{\sigma^2_w} \distas{} \N \bigl(0,1\bigr)$.

%%%%%%%%%%%%%%%%%%%%%%%%%%%%%%%%
\noindent When $\gamma=2$, the mean and variance of $\W$ are as follows
\begin{align}
\E(\W^2) %
& = \E\Biggl[\Biggl(\beta_0+  \sum_{j=1}^J \beta_j x_{it,j}\Biggr)^2\Biggr]\notag \\
& = \beta^2_0+  \sum_{j=1}^J \beta^2_j \, \E\bigl( x^2_{it,j}\bigr)+ 2 \beta_0 \sum_{j=1}^J \beta_j \, \E\bigl( x_{it,j}\bigr) + 2 \sum_{\substack{j,k=1\\j \ne k}}^J\beta_j\beta_k  \E\bigl( x_{it,j}\bigr) \E\bigl( x_{it,j}\bigr) \label{eq:mean_w2_iid}\\
& = \beta^2_0+  \sum_{j=1}^J \beta^2_j (\sigma^2_{x_j}+\mu^2_{x_j})+  2 \beta_0 \sum_{j=1}^J \beta_j \mu_{x_j}+ 2\sum_{\substack{j,k=1\\j \ne k}}^J\beta_j\beta_k \mu_{x_j}\mu_{x_j}\notag \\
%%%%
\mathrm{Var}\bigl(\W^2\bigl) %
& = \mathrm{Var}\Biggl[\biggl(\beta_0+  \sum_{j=1}^J \beta_j x_{it,j}\biggr)^2\Biggr]\notag \\
& =   \sum_{j=1}^J \beta^4_k \, \mathrm{Var}\bigl( x^2_{it,j}\bigr)+ 4 \beta^2_0 \sum_{j=1}^J \beta^2_j \, \mathrm{Var}\bigl( x_{it,j}\bigr)+ 4 \sum_{\substack{j,k=1\\j \ne k}}^J\beta^2_j\beta^2_k  \mathrm{Var}\bigl( x_{it,j}, x_{it,j} \bigr)\notag\\
& =   \sum_{j=1}^J \beta^4_k  \sigma^2_{x_j}+ 4 \beta^2_0 \sum_{j=1}^J \beta^2_j \sigma^2_{x_j}+ 4 \sum_{\substack{j,k=1\\j \ne k}}^J\beta^2_j\beta^2_k  \bigl(\sigma^2_{x_j} \sigma^2_{x_j} +  \sigma^2_{x_j}\mu^2_{x_j}+ \sigma^2_{x_j}\mu^2_{x_j}\bigr) \label{eq:var_w2_iid}
\end{align}
\noindent  where $\E\bigl( x^2_{it,j}\bigr)=\sigma^2_{x_j}+\mu^2_{x_j}$, $\E\bigl( x_{it,j}, x_{it,k}\bigr)=\E\bigl( x_{it,j}\bigr) \E\bigl( x_{it,k}\bigr)=\mu_{x_j}\mu_{x_k}$,  and  $\mathrm{Var}\bigl( x_{it,j}, x_{it,k} \bigr) =\bigl[\E\bigl( x_{it,j}, x_{it,ks}\bigr)\bigr]^2-\E\bigl( x_{it,j}\bigr)^2 \E\bigl( x_{it,ks}\bigr)^2 = \sigma^2_{x_j} \sigma^2_{x_ks} +  \sigma^2_{x_j}\mu^2_{x_k}+ \sigma^2_{x_j}\mu^2_{x_k}$ because of the  \emph{iid} assumption. Any covariance among variables is null because of the assumption of independence. With standardised $\W$, $\Bigr(\frac{\W-\mu_w}{\sigma^2_w}\Bigr)^2 \distas{}  \chi^2_1$.

 %$\mathrm{Var}\bigl( x_{it,j} x_{it,j} \bigr) = \E\bigl[ x^2_{it,j} x^2_{it,j}\bigr] - \E\bigl[ x_{it,j}x_{it,j}\bigr]^2 =  \E\bigl( x^2_{it,j}\bigr) \E\bigl[ x^2_{it,j}\bigr]  - \E\bigl( x_{it,j}\bigr)^2 \E\bigl( x_{it,j}\bigr) ^2 = (\sigma^2_{x_j}+\mu^2_{x_j})( \sigma^2_{x_j}+\mu^2_{x_j}) - \mu^2_{x_j}  \mu^2_{x_j}=\sigma^2_{x_j} \sigma^2_{x_j} +  \sigma^2_{x_j}\mu^2_{x_j}+ \sigma^2_{x_j}\mu^2_{x_j}$.

%%%%%%%%%%%%%%%%%%%%%%%%%%
\end{document}